\documentclass[prx,  twocolumn, showpacs, floatfix, tightenlines, amsmath, amssymb, superscriptaddress, longbibliography]{revtex4-2}

\usepackage{overpic}
\usepackage{graphicx}
\usepackage{color}
\usepackage{dcolumn}
\usepackage{hyperref}
\usepackage[nolist,nohyperlinks]{acronym}
\usepackage{CJK}
\usepackage{braket}
\usepackage{ragged2e}

\usepackage{amsmath}
\usepackage[varg]{txfonts}
\usepackage[utf8]{inputenc}
\usepackage{times}
\usepackage{mathrsfs}
\usepackage{bm}
\usepackage{bibunits}
\usepackage[T1]{fontenc}

\defaultbibliography{arXiv}
\defaultbibliographystyle{apsrev4-2}

\definecolor{cblue}{RGB}{19,107,192}
\hypersetup{colorlinks=true,linkcolor=cblue,citecolor=cblue,urlcolor=cblue}
\newenvironment{subalign}{\subequations\align}{\endalign\endsubequations}

\newcommand{\subref}[2]{\ref{#1}\hyperref[#1]{#2}}

\newcommand{\nocontentsline}[3]{}
\newcommand{\tocless}[2]{\bgroup\let\addcontentsline=\nocontentsline#1{#2}\egroup}

\newcommand{\degree}{\ensuremath{^{\circ}}\,\,}

\newcommand{\tsup}[1]{\textsuperscript{#1}}
\newcommand{\tsub}[1]{\textsubscript{#1}}
\newcommand{\avg}[1]{\braket{#1}}

\newcommand{\del} {\partial}
\newcommand{\h}[1]{{#1}^{\dagger}}
\newcommand{\nh}[1]{{#1}^{\phantom \dagger}}

\newcommand{\cb}[1]{\bar{#1}}

\newcommand{\hc}{\textrm{h.c.}}

\renewcommand{\vec}[1]{\boldsymbol{#1}}
\newcommand{\mat}[1]{{\vec{#1}}}
\newcommand{\trp}[1]{{#1}^{\intercal}}
\newcommand{\vhat}[1]{\vec{\hat{#1}}}

\newcommand{\meV}{\ \textrm{meV}}

\newcommand{\KIC}{K\tsub{2}IrCl\tsub{6}}
\newcommand{\SrIrO}{Sr\tsub{2}IrO\tsub{4}}

\begin{document}

\title{Pulling order back from the brink of disorder: Observation of a nodal line spin liquid and fluctuation stabilized order in \KIC{} }

\author{Qiaochu Wang}
\affiliation{Department of Physics, Brown University, Providence, Rhode Island 02912, United States}

\author{Alberto de la Torre}
\thanks{Current address: Department of Physics, Northeastern University, Boston, MA 02115}
\affiliation{Department of Physics, Brown University, Providence, Rhode Island 02912, United States}

\author{Jose A. Rodriguez-Rivera}
\affiliation{NIST Center for Neutron Research, National Institute of Standards
    and Technology, Gaithersburg, MD 20899, USA}
\affiliation{Department of Materials Science and Engineering, University of
    Maryland, College Park, MD 20742, USA}
 
\author{Andrey A. Podlesnyak}
\author{Wei Tian}
\author{Adam A. Aczel}
\author{Masaaki Matsuda}
\affiliation{Neutron Scattering Division, Oak Ridge National Laboratory, Oak Ridge, Tennessee 37831, USA}

\author{Philip J. Ryan}
\author{Jong-Woo Kim}
\affiliation{X-ray Science Division, Argonne National Laboratory, Argonne, Illinois 60439, USA}

\author{Jeffrey G. Rau}
\affiliation{Department of Physics, University of Windsor, Windsor, Ontario, N9B 3P4, Canada}

\author{Kemp W. Plumb}
\email[Corresponding author: ]{kemp\_plumb@brown.edu}	
\affiliation{Department of Physics, Brown University, Providence, Rhode Island 02912, United States}

\date{\today}
\begin{abstract}
Competing interactions in frustrated magnets can give rise to highly degenerate ground states from which correlated liquid-like states of matter often emerge. The scaling of this degeneracy influences the ultimate ground state, with extensive degeneracies potentially yielding quantum spin liquids, while sub-extensive or smaller degeneracies yield static orders. A longstanding problem is to understand how ordered states precipitate from this degenerate manifold and what echoes of the degeneracy survive ordering. Here, we use neutron scattering to experimentally demonstrate a new ``nodal line'' spin liquid, where spins collectively fluctuate within a sub-extensive manifold spanning one-dimensional lines in reciprocal space. Realized in the spin-orbit coupled, face-centered cubic iridate \KIC{}, we show that the sub-extensive degeneracy is robust, but remains susceptible to fluctuations or longer range interactions which cooperate to select a magnetic order at low temperatures. Proximity to the nodal line spin liquid influences the ordered state, enhancing the effects of quantum fluctuations and stabilizing it through the opening of a large spin-wave gap. Our results demonstrate quantum fluctuations can act counter-intuitively in frustrated materials: instead of destabilizing ordering, at the brink of the nodal spin liquid they can act to stabilize it and dictate its low-energy physics.
\end{abstract}

\maketitle

\acrodef{HK}[HK]{Heisenberg-Kitaev}
\acrodef{FCC}[FCC]{face-centered cubic}
\acrodef{LSWT}[LSWT]{linear spin wave theory}
\acrodef{NLSWT}[NLSWT]{non-linear spin wave theory}
\acrodef{NNN}[NNN]{next nearest-neighbor}
\acrodef{AFM}[AFM]{antiferromagnet}
\begin{bibunit}
The promise of discovering novel states of matter governed by quantum and thermal fluctuations drives a sustained interest in highly frustrated magnetic materials. One of the oldest and most important examples is the Heisenberg \ac{AFM} on the \ac{FCC} lattice. Anderson~\cite{anderson:1950} showed that classically its ground state degeneracy scales sub-extensively in the system size, and includes co-planar spin spirals characterized by all magnetic wavevectors spanning one-dimensional lines in reciprocal space. As temperature is lowered, fluctuations of the spins become more confined to states on these lines, forming a correlated state we term a ``nodal line spin liquid''. This spin-liquid  is distinguished from a conventional paramagnet by having growing correlations along two spatial directions as temperature is lowered, but only short-range correlations along the third. Among correlated paramagnets, including spiral spin-liquids, where the degenerate states span surfaces in momentum space~\cite{bergman:2007, gao:2017}, or classical spin liquids~\cite{villain:1979, reimers:1991, Moessner:1998, henley:2010, plumb:2019}, 
where degenerate states span a volume, the nodal line spin liquid has the least degeneracy, and thus sits closest to conventional magnetic order. 

Due to its sub-extensive degeneracy, the nodal line spin liquid is susceptible to order via quantum or thermal fluctuations -- ``order-by-disorder"~\cite{Villain:1980, henley:1987, rastelli:1987, henley:1989}. For the \ac{FCC} Heisenberg \ac{AFM}, thermal~\cite{gvozdikova:2005, zhitomirsky:2012} and quantum~\cite{schick:2020} fluctuations each favor a sub-extensive set of collinear N\'eel states~\cite{anderson:1950, danielian:1960}, both ultimately selecting so-called ``type I'' magnetic order characterized by a $(1,0,0)$ propagation vector. However, the selection of this type I order among the collinear states is extremely weak -- with an energy difference of only $\sim\!0.25\%$ of the Heisenberg exchange, $J$, and many other metastable collinear states lying within $\sim\! 1\%$ of $J$ in energy~\cite{schick:2020}. 
While this classical degeneracy and fluctuation-induced selection are robust to nearest-neighbor anisotropic exchange interactions, it can be lifted energetically by further neighbor couplings. 
For example, \ac{NNN} Heisenberg exchange selects ``type I'' order when ferromagnetic, but selects the ``type III'' state, a collinear N\'eel order with propagation vector $(1,0.5,0)$ when antiferromagnetic~\cite{yildirim:1998}.

\begin{figure*}[t!]
    \centering
    \includegraphics[]{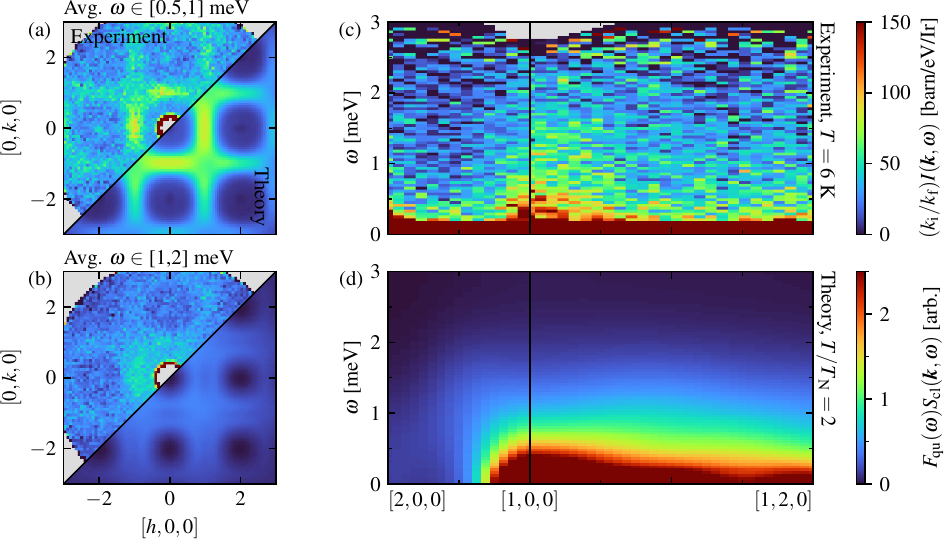}
    \caption{\label{fig:para} Dynamic correlations of the nodal line spin liquid. (a),  (b), Constant energy neutron scattering intensity at $T=6$~K integrated over 0.5 to 1~meV in (a) and 1 to 2~meV in (b). Lower right of (a), and  (b), shows the result from classical spin dynamics of our best fit  FCC Heisenberg-Kitaev model with exchange parameters $J=0.74$~meV and $K=0.15$~meV.
    (c), Momentum and energy resolved magnetic excitation spectra of the nodal line spin liquid at $T=6$~K. Momentum directions are integrated over $\pm 0.15$~r.l.u. . (d), Classical spin dynamics at a temperature twice the model's ordering temperature computed as described in~\cite{Supplemental}.}
\end{figure*}

In each case, ordering out of the nodal-line-spin-liquid proceeds through a multi-step process. First, order-by-disorder singles out the sub-manifold of discrete collinear states from the continuous manifold of the nodal line spin liquid.  Despite their near degeneracy, each of these local minima exhibits conventional magnetic order with gapped spin-wave excitations. Second, a specific collinear state among the local minima is selected by either fluctuations or energetic effects. 
While the ordered state is ultimately determined by the second stage, the low-energy physics -- including the excitations and thermodynamic properties -- are dictated by the fluctuation-induced selection arising in the first stage. 
This leads to a counter-intuitive result: while the final order-by-disorder selection of the ground state is very weak, each of the collinear states is stable, with a large spin-wave gap generated by strong quantum fluctuations along the nodal line directions of the parent spin liquid. 
This large gap then suppresses fluctuations of the ordered moment. Thus, despite the difficulty of fluctuations in selecting a unique ground state, they act to stabilize whichever particular ordering is singled out by sub-leading perturbations, and ultimately dictates the physics of the material.

In this article, we show that such physics is realized in the \ac{FCC} iridate \KIC{}. 
First, we establish that the magnetic correlations form one-dimensional rods in reciprocal space characteristic of a nodal line spin liquid as the magnetic transition is approached from above. 
We then present a comprehensive series of neutron scattering measurements in the ordered state that, when combined with theoretical modeling of the dynamic spin correlations, conclusively demonstrate that \KIC{} is a nearly ideal realization of the Heisenberg-Kitaev model on the \ac{FCC} lattice. 
This model is highly frustrated, with a specific ordering only weakly selected by \ac{NNN} interactions or quantum fluctuations. 
Despite the appearance of static magnetic order, as observed by diffraction measurements, a hard excitation gap, and long-lived spin waves, \ac{LSWT} theory completely fails as a description of magnetic excitations in \KIC{}. 
Instead, we find that the physics of the excitation spectrum is dominated by fluctuation effects that generate a large, $\sim 0.7$~meV, excitation gap -- about 30~\% of the overall $\sim 2.5$~meV magnon bandwidth. 
We thus show how proximity to the nodal line spin liquid enhances quantum fluctuations and, contrary to conventional lore, stabilizes magnetic order by quenching out the zero-point corrections to the ordered moment.

While much theoretical and experimental effort has been made on \ac{FCC} magnets ~\cite{bailey:1959,   haar:1962, haar:1962a, lines:1963, tahir-kheli:1966, matsuura:2003, aczel:2016, balla:2020}, the antifluorite-type \KIC{} has sustained  more than 70 years of interest~\cite{owen:1959, bailey:1959, griffiths:1959, cooke:1959, judd:1959, haar:1962, haar:1962a, lines:1963, harris:1965, tahir-kheli:1966, hutchings:1967, minkiewicz:1968, lynn:1976, willemsen:1977, moses:1979, khan:2019, reig-i-plessis:2020, bhaskaran:2021}. 
Despite such effort, the magnetic Hamiltonian of \KIC{} has yet to be resolved due to an absence of any inelastic neutron scattering data. 
In addition to the presence of geometric frustration, the magnetic degrees of freedom in \KIC{} are spin-orbital entangled $j=1/2$ states~\cite{jackeli:2009}, supporting bond anisotropic Kitaev interactions, which are believed to be significant relative to the Heisenberg exchange~\cite{cooke:1959, khan:2019}. 
Early magnetic neutron diffraction experiments revealed a type III antiferromagnetic order below $\sim\! 3$~K~\cite{hutchings:1967}. 
Based on this observation and a consistent explanation of the bulk magnetic properties, \KIC{} has long been considered to represent a near ideal \ac{FCC} lattice with weak antiferromagnetic \ac{NNN} interactions~\cite{khan:2019, reig-i-plessis:2020}. 
However, recent electron spin resonance measurements indicate \ac{NNN} terms may only play a perturbative role and order-by-disorder physics may be essential~\cite{bhaskaran:2021} in explaining its properties. 
To date, there are no reported measurements of momentum resolved dynamic correlations characterizing the excitation spectrum of this highly frustrated anisotropic magnet.

\tocless{\section{Nodal line spin liquid}}
We first present inelastic neutron scattering results that  demonstrate two-dimensional long-range correlations above the N\'eel transition at $T=6$~K, well above our observed ordering temperature $T_{\rm N}=3.2$~K, through characteristic rods of intensity in the dynamical structure factor averaged over a broad energy window [Fig.~\ref{fig:para}]. 
Such highly anisotropic, two dimensional, correlations arising from an underlying isotropic, three dimensional, magnetic Hamiltonian are the direct experimental signature of a nodal-line spin liquid.

The rods of intensity extend along equivalent $[1,q,0]$ reciprocal space directions characteristic of the set of co-planar spin spirals of the nodal line spin liquid. Momentum-energy slices shown in Fig.~\subref{fig:para}{(c)}  reveal a continuum of scattering extending from these rods, with intensity concentrated near $\vec{Q}=(1,0,0)$ momentum transfers and extending to 2~meV. 
There is no evidence for a build-up of critical intensity near the $\vec{Q}=(1,0.5,0)$ wavevector of the type III magnetic ordering that appears below $T_{\rm N}$. 
Instead, dynamic magnetic correlations are localized along lines in reciprocal space, identifying \KIC{} as a \emph{nodal-line spin-liquid} state in this temperature regime. 

We compare these correlations with expectations from the classical spin dynamics of the \ac{FCC} \ac{HK} model.
As shown in Fig.~\subref{fig:para}{(a)}, the low energy momentum dependent correlations we observe are well captured by classical spin dynamics of the \ac{FCC} \ac{HK} model within its paramagnetic phase that we will further establish below as the appropriate Hamiltonian for \KIC{}. 
Although the classical calculation can reproduce much of the observed momentum dependent scattering, there is a notable discrepancy for energy transfers above 1~meV where the data reveals an intensity maxima near $\vec{Q}=(1,0,0)$ not seen in the theoretical calculation [Fig.~\subref{fig:para}{(b,d)}]. 
This discrepancy reflects strong quantum fluctuations in \KIC{} that are not captured in the classical spin dynamics and redistributes magnetic fluctuations to higher energies. 

Since the degeneracy of the nodal line spin liquid scales subextensively, either fluctuations, small additional magnetic interactions, or both are expected to select a long range ordered state as temperature is lowered. 
Consistent with these expectations, \KIC{} undergoes a magnetic ordering transition at $T_{\rm N}=3.2$~K. 
However, we will show below that, despite a well formed type III N\'eel order, the excitations out of the ordered state cannot be captured on even a qualitative level by semi-classical \ac{LSWT} starting from the ordered state and, as in the paramagnetic phase, large corrections from quantum fluctuations are required to capture the physics of this material.

\begin{figure*}[t]
    \centering
    \includegraphics[width=\textwidth]{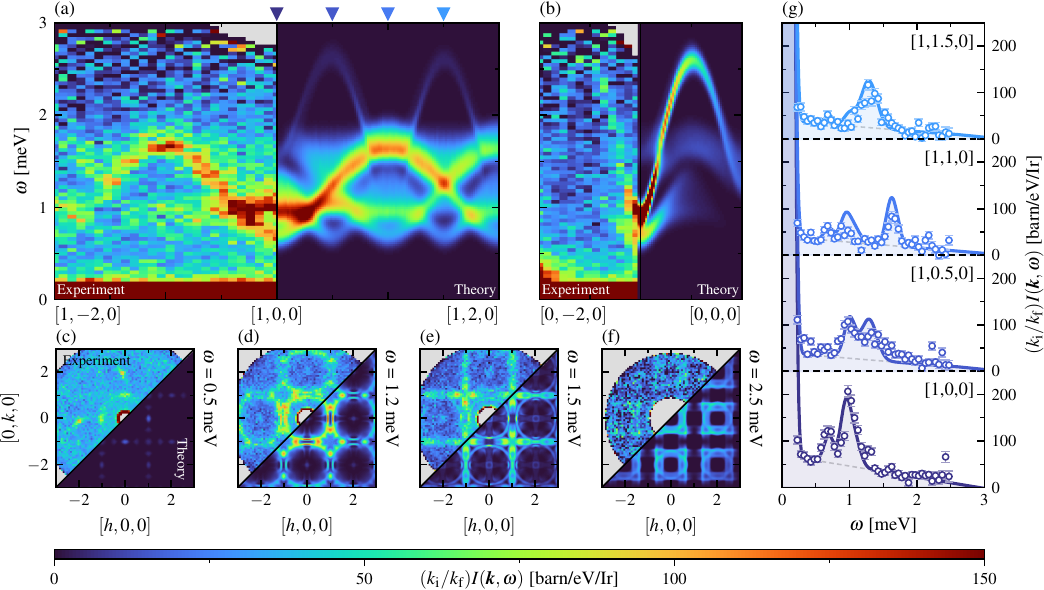}
    \caption{\label{fig:spin-waves}Spin waves in the ordered phase.  Measured (left) and calculated (right) energy–momentum slices along (a), $[1,k,0]$ and  (b), $[0,k,0]$ integrated over $h \pm 0.15$~r.l.u. and $l \pm 0.15 $~r.l.u..  (c)-(f),  Constant energy slices across the magnon spectra integrated over $ \omega \pm 0.15$~meV. The measured spectrum is shown in the upper left and the calculated spectrum is shown in the lower right. (g), Constant $\vec{Q}$ cuts at high symmetry wavevectors indicated by colored arrows at the top of (a) and integrated over $\pm~0.1$~r.l.u.. Data in panels (a)--(g) was collected at $T=300$~mK and errorbars represent one standard deviation. Spectrum calculated at $O(1/S^2)$ using self-consistent non-linear spin-wave theory for an FCC Heisenberg-Kitaev model with type III magnetic order (domain averaged) using best-fit exchange parameters $J=0.74$~meV and $K=0.15$~meV.}
\end{figure*}

\tocless{\section{Fluctuation induced spin wave gap}}
Fig.~\ref{fig:spin-waves} shows an overview of the magnetic neutron spectra at $T=300$~mK, well below the onset of type III $(0.5, 0, 1)$ magnetic order. 
There are sharp spin waves present throughout the Brillouin zone spanning a bandwidth of $2.5$~meV with a $0.7 \pm 0.05$~meV gap at the $(1,0,0)$ dispersion minimum. 
Although the strong neutron absorption from Ir places limitations on the  signal-to-noise and achievable energy resolution of our experiments, the spectrum we observe is dominated by resolution limited single magnon excitations, and any multi-magnon scattering was not measurable. 
The observed spectrum is surprising for two reasons. 
First, the dispersion minimum does not occur at the wavevector of static magnetic order $(1,0.5,0)$, but instead near $(1,0,0)$, and second for the generic symmetry allowed spin model relevant to \KIC{} \ac{LSWT} predicts a gapless pseudo-Goldstone mode in the type III phase~\cite{rau:2018,Supplemental}. 

The presence of a large excitation gap could potentially be explained within \ac{LSWT} by the lowering of the crystal symmetry from cubic to (say) tetragonal, as might arise from a weak structural distortion at $T_{\rm N}$~\cite{aczel:2016, wang:2024}. 
Although all previous diffraction measurements have reported \KIC{} as cubic for $T<T_{\rm N}$~\cite{khan:2019, reig-i-plessis:2020}, there is spectroscopic evidence that \KIC{} may exhibit local structural distortions~\cite{lee:2022}. 
Furthermore, recent high-resolution single crystal x-ray diffraction measurements reveal a 0.02\% compression of the c-axis accompanied by a 0.014\% expansion of $a$ and $b$ axes coinciding with the magnetic transition~\cite{wang:2024}. 
To conclusively identify this large spin-wave gap as fluctuation induced, we first determine whether these distortions could potentially account for it's observed magnitude.

Since $j=1/2$ pseudospins have both a spin and orbital contribution to their magnetic moment, they can couple more easily with lattice deformations of the appropriate symmetry~\cite{liu:2019}. 
Due to this coupling, the type III ordering will necessarily generate structural deformations through a pseudo-Jahn-Teller mechanism, which is consistent with the observed tetragonal distortion~\cite{wang:2024} and the reported anisotropic magnetization density in \KIC{}~\cite{lynn:1978}. 
Such magneto-elastic couplings will also give rise to spatially anisotropic exchange interactions, $D$, that can induce a magnon gap $\propto \sqrt{JD}$~\cite{liu:2019}, where $J$ is the (dominant) Heisenberg exchange. 
We estimate the magnetic anisotropy from the measured tetragonal strain $\varepsilon \approx 2\times10^{-4}$~\cite{wang:2024} and spin-lattice coupling constant $\tilde{g}\simeq\left(t^2/U\right)3/2\lambda\left(J_{\rm H}/U\right)$ where $t$, $U$, $J_{\rm H}$, and $\lambda$ are the hopping amplitude, Coulomb repulsion, Hund's coupling, and spin-orbit coupling, respectively~\cite{liu:2019}. 
Based on similar estimates in \SrIrO{}~\cite{liu:2019}, but with a nearest-neighbor exchange constant $J\sim t^2/U$ in \KIC{} that is a factor of $\sim 100$ smaller, we estimate $\tilde{g}\approx0.25$~meV and thus $D \sim \tilde{g} \varepsilon \approx 0.025\ \mu$eV. 
This yields a distortion-induced gap $\sim 0.05$~$\mu$eV -- \emph{four-orders of magnitude too small} to account for the observed  0.7~meV value~\cite{Supplemental}. 
As an additional check, we have carried out a \ac{LSWT} calculation including all symmetry-allowed terms with only tetragonal symmetry~\cite{Supplemental}. 
This includes different exchanges within ($J$, $K$) and perpendicular to  ($J_{\perp}, K_{\perp}$) the basal plane (perpendicular to the tetragonal axis). 
Fitting a tetragonal model that can account for the measured gap yields significant differences in the exchanges within the plane and in the perpendicular direction; the Heisenberg exchange yielding a modest $J_{\perp}/J = 1.29$, but a much larger $K'_{\perp}/K = 1.93$ for the Kitaev exchange~\cite{Supplemental}.
Given the smallness of the observed tetragonal distortion and our estimate of the magneto-elastic coupling, we find this level of spatial anisotropy in the exchanges implausible and rule it out as an explanation for the large spin-wave gap in our data.

Given that magnetic excitations in the ordered state of \KIC{} cannot be explained qualitatively by \ac{LSWT} when cubic symmetry is maintained, we must then consider the effects of quantum fluctuations beyond \ac{LSWT}.  
Including interactions between the spin-waves can lift the pseudo-Goldstone modes, inducing a gap through quantum fluctuations~\cite{schick:2022,rau:2018}. 
To account for these fluctuation induced gaps, we have computed the dynamical structure factor for the \ac{FCC} \ac{HK} model (as measured by INS) about the type III ordering using $O(1/S^2)$ self-consistent \ac{NLSWT}~\cite{Supplemental}. 
As shown in Fig.~\ref{fig:spin-waves}, we find excellent quantitative agreement with our experimental data -- including the large gap -- using only nearest-neighbor Heisenberg, $J=0.74$ meV, and Kitaev, $K=0.15$ meV ($K/J=0.2$), exchanges. 
These exchange parameters are comparable to DFT predictions~\cite{khan:2019} and estimates from ESR measurements~\cite{bhaskaran:2021}. 
Neither the symmetric anisotropy $\Gamma$ nor \ac{NNN} interactions are required to describe the inelastic neutron data for \KIC{}, suggesting they are small relative to $J$ and $K$. 

However, despite providing a quantitative description of the excitations in \KIC{}, without further neighbor interactions our best-fit model does not clearly predict a type III ground state. Even including fluctuations to $O(1/S^2)$ the energy difference between type I and III states is expected to be incredibly small~\cite{schick:2022}. 
This close competition between type I and type III states indicates the mechanism selecting type III order in \KIC{} is delicate and the ordering transition in \KIC{} deserves closer scrutiny.

\begin{figure*}[t]
    \centering
    \includegraphics[]{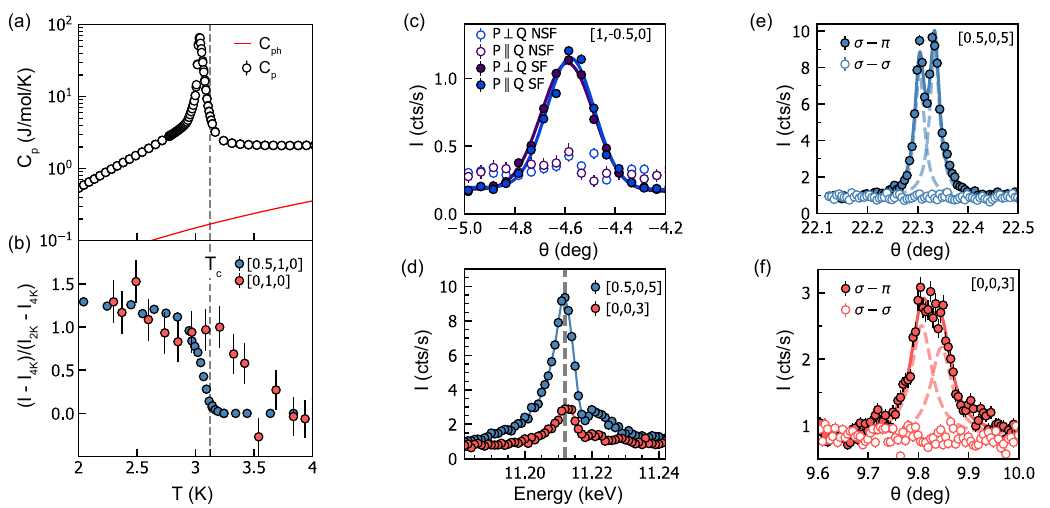}
    \caption{\label{fig:order} Coexistence of type I and type III collinear magnetic orders in \KIC{}. (a), Measured zero-field specific heat $C_p$ and estimated lattice contribution $C_{\rm ph}$. (b), Temperature dependence of $(0.5,1,0)$ (type III) and $(0,1,0)$ (type I) intensities obtained from neutron diffraction. (c), Polarized neutron diffraction of a type III peak at 1.5~K indicating the absence of spin component along the $[0,0,1]$ direction. (d), Energy dependent elastic X-ray scattering demonstrating that both peaks arise from a magnetic resonance. Resonant x-ray rocking curves on (e), type III  and (f), type I peaks at 2~K measured on the same crystal as for (a). Errorbars in all panels represent one standard deviation.}
\end{figure*}

\tocless{\section{Competing type I and III magnetic order}}
Heat capacity measurements on our samples display a single sharp peak at $3.2$~K, in agreement with the critical temperature extracted from the $(1,0.5,0)$ magnetic order parameter [Fig.~\subref{fig:order}{(a,b)}]. 
However, a broader reciprocal space survey on the same samples revealed additional weak magnetic Bragg reflections at $\vec{Q}=(1,0,0)$, corresponding to type I order, with an onset temperature of $T_{\rm N}^{\rm I}=3.4$ K [Fig.~\subref{fig:order}{(b)}]. 
The magnetic origin of $(1,0,0)$ reflections was confirmed through energy and polarization-dependent resonant x-ray diffraction that indicates an absence of any charge signal [Fig.~\subref{fig:order}{(d-f)}]. 
The fine momentum space resolution x-ray measurements also revealed that $(0.5, 0 , 5)$ magnetic Bragg peaks, corresponding to type III N\'eel order, are resolution limited, while the $(0,0,3)$ magnetic peak, corresponding to type I order, has a finite correlation length of  877~$\textrm{\AA}$. 
Although there is no heat capacity anomaly visible at $T_{\rm N}^{\rm I}$, we find substantial magnetic entropy extending up to $T \sim 15$~K that is consistent with the onset of short-range type I order at $3.5$~K.

 A simultaneous appearance of $(1,0.5,0)$ and $(1,0,0)$ magnetic peaks is consistent with either  a coexistence of collinear magnetic phases across multiple domains, or a  multi-$Q$ order.  Polarized neutron diffraction presented in Fig.~\subref{fig:order}{(c)} rules out the possibility of non-coplanar multi-$Q$ states. 
 The spin-flip neutron intensity of the $\vec{Q}=(1,-0.5,0)$ Bragg reflection is sensitive to both in and out-of-plane components of the magnetization for guide fields $\vec{P}$ parallel to $\vec{Q}$, and in-plane components of the magnetization, along $[1,2,0]$, for $\vec{P}\perp \vec{Q}$. 
 While the non-spin flip neutron intensity is only sensitive to out-of-plane components of the magnetization, along $[0,0,1]$, when $\vec{P}\perp \vec{Q}$. 
 We observed identical $(1, -0.5, 0)$ peak intensities for $ \vec{P} \parallel \vec{Q}$ and $\vec{P} \perp \vec{Q}$ spin-flip channels  and no magnetic signal in the non-spin flip component, ruling out any out-of-plane component, as expected for a collinear type III N\'eel state [Fig.~\subref{fig:order}{(c)}]. 
 We also find that the polarized diffraction from $\vec{Q}=(1,0,0)$ magnetic peaks is most consistent with a collinear arrangement of magnetic moments that are parallel to the $[0,0,1]$ direction, expected for a type I N\'eel state~\cite{Supplemental}. 
 Thus, the diffraction data support a coexistence of type I and type III magnetic orders across different spatial regions of the sample. This is consistent with expectations from order-by-disorder, which strongly favor collinear single-$Q$ states. From the integrated neutron diffraction Bragg intensities, we extract a relative phase fraction $N_{\rm I}/N_{\rm III}\approx 11\%$. 
 This phase fraction did not change between measurements on two different samples from different crystal growths indicating that the small fraction of type I phase is likely intrinsic and does not arise from quenched disorder in our samples. 

The appearance of both type I and III order near $T_{\rm N}$ in \KIC{} implicates weak and competing selection effects from fluctuations or further-neighbor exchange for these nearly degenerate states. To better understand the role of these selection mechanisms near $T_{\rm N}$, we consider energetic selection via a \ac{NNN} exchange $J_2$ in the classical model. When $J_2/J=0$, we find that thermal order by disorder selects a type I order for the \ac{FCC} \ac{HK} model~\cite{gvozdikova:2005, schick:2020, Supplemental}. A small $J_2/J \sim 0.02$ selects type III order instead, onsetting at a nearly identical N\'eel temperature to the $J_2=0$ case~\cite{Supplemental}.  Together, our data and modeling demonstrate that \KIC{} uniquely displays a competition between energetic and entropic mechanisms for order selection. Thermal fluctuations initially favor type I order, but as temperature is lowered and the influence of entropy diminishes, quantum fluctuations or weak further neighbor interactions select type III order. Such a competition between weak selection mechanisms for the magnetic order in \KIC{} is consistent with nearly negligible \ac{NNN} interactions and underscores the dominant role played by quantum fluctuations in this material.

\tocless{\section{Discussion}}
The conventional lore of frustrated quantum magnetism is that  degenerate and competing ground states suppress the tendency towards magnetic long range order. 
In the quantum limit of $S=1/2$, quantum fluctuations among degenerate ground states persist as $T\rightarrow 0$ and act to reduce the static ordered moment size. 
Experimentally, this manifests as a transfer of neutron spectral weight from the magnetic Bragg peak (order parameter) to the excitations~\cite{Lee:1997}. 
\KIC{} presents an important counterpoint to these expectations where despite the appearance of a well formed static N\'eel state and long lived magnon excitations, higher order (in $1/S$) corrections from quantum fluctuations are essential to describe the spectra on even a qualitative level. 
The ordered moment reduction can be quantified through the energy distribution of neutron intensity. The momentum and energy integrated intensity gives the total moment $S(S+1)$, while the elastic intensity for a fully ordered moment is $S^2$. 
Longitudinal fluctuations will reduce the static moment so that elastic intensity is $(S-\Delta S)^2$ and the ratio of elastic to total integrated intensity is $R=(S- \Delta S)^2/S(S+1)$. 
For \KIC{}, we find $R \approx 0.31$, or $\Delta S \rightarrow 0$, which is consistent with the small moment reduction predicted by \ac{NLSWT} and demonstrates that by generating a large gap for longitudinal magnetic fluctuations, quantum fluctuations in \KIC{} act to stabilize the ordered moment.

Above $T_{\rm N}$, \KIC{} is a unique realization of a new classical spin liquid: the nodal line spin liquid. 
This correlated paramagnet is similar to the spiral-spin liquid recently observed in MnSc$_2$S$_4$~\cite{gao:2017}. 
Both have a subextensive degeneracy and  are susceptible to thermal order-by-disorder, but fluctuations in the nodal line liquid are confined to one dimensional lines instead of spiral surfaces.  Despite this reduced degeneracy, the role of quantum fluctuations is significantly enhanced in \KIC{}. 
Magnetic ordering in MnSc$_2$S$_4$ is controlled by further neighbor, dipolar, and anisotropic exchange interactions, the excitation spectrum is fully described by \ac{LSWT}~\cite{krimmel:2006, gao:2020}, and there is no experimental support for an order by disorder transition~\cite{krimmel:2006, gao:2018, gao:2020}. 
This difference likely arises from the small $j=1/2$ moments (Mn$^{2+}$ is $S=5/2$) and negligible dipolar interactions in \KIC{}. 

It is the scale of quantum fluctuations in \KIC{} that distinguish it as an important frustrated magnet. 
These  can be quantified  by the relative size of the fluctuation induced magnon gap compared to the bandwidth that exceeds 30 \% in \KIC{}. 
For comparison, quantum order-by-disorder  spin wave gaps have been found in the garnet Fe$_2$Ca$_3$(GeO$_4$)$_3$~\cite{brueckel:1988}, the pyrochlore Er$_2$Ti$_2$O$_7$~\cite{champion:2003, ruff:2008, savary:2012, ross:2014}, and the honeycomb lattice CoTiO$_3$~\cite{yuan:2020, elliot:2021}. 
In all of these, the spin wave gap to bandwidth ratio is less than 10\% and nearly all details of the magnon spectra besides the fluctuation induced gaps are reproduced by linear spin wave theory, while this fails qualitatively for \KIC{}.

The scale of quantum fluctuations in \KIC{} is further reflected by the dynamics of the nodal line spin liquid phase that are only partially reproduced by a classical spin dynamics calculation. 
This is distinct from other correlated paramagnets. For example even the excitation spectrum of the highly frustrated Heisenberg model on the pyrochlore lattice can be qualitatively captured by classical spin dynamics calculations renormalized by a classical to quantum correspondence factor~\cite{zhang:2019}. 
Departures from renormalized classical spin dynamics calculations in the nodal line spin liquid realized by \KIC{} calls for further work to understand the quantum to classical crossover in classical spin liquids.
\tocless{\section*{Acknowledgments}}
Work at Brown University was supported by the U.S.  Department of Energy, Office of Basic Energy Sciences, under Grant No. DE-SC0021223. A portion of this research used resources at the Spallation Neutron Source and High Flux Isotope Reactor, the DOE Office of Science User Facilities operated by the Oak Ridge  National Laboratory. Access to MACS was provided by the Center for High Resolution Neutron Scattering, a partnership between the National Institute of  Standards and Technology and the National Science Foundation under Agreement  No.  DMR-2010792. This research used resources of the Advanced Photon Source, a U.S. Department of Energy (DOE) Office of Science user facility operated for the DOE Office of Science by Argonne National Laboratory under Contract No. DE-AC02-06CH11357. Work at the University of Windsor was funded by the Natural Sciences and Engineering Research Council of Canada (NSERC) (Funding Reference No. RGPIN-2020-04970). We acknowledge the use of computational resources provided by Digital Research Alliance of Canada.
\appendix
\setcounter{equation}{0}
\renewcommand{\theequation}{A\arabic{equation}}

\tocless{\section{Methods}}
\noindent
\tocless{\subsection{Sample Synthesis and Characterization}}
Commercially available powder of \KIC{} was obtained from Fisher Scientific. \KIC{} crystals were prepared by slow evaporation from a saturated solution of \KIC{} in dilute hydrochloric acid (pH~$\sim 3$). The evaporation process was controlled to maintain a temperature of 32 degrees Celsius. Larger crystals were obtained by successive seeding of solution growths.

Specific heat and magnetization were measured in a Quantum Design Physical Properties Measurement System (PPMS). Specific heat was measured upon warming in zero magnetic field after a zero-field cooling.  Additional sample characterization is provided in the Supplemental information~\cite{Supplemental}.

\tocless{\subsection{Neutron Scattering}}
Inelastic neutron scattering measurements were performed on a sample comprised of 11 co-aligned crystals with total mass 0.3~g using the Cold Neutron Chopper Spectrometer (CNCS) at the Spallation Neutron Source (SNS) at Oak Ridge National Laboratory (ORNL) and the MACS spectrometer~\cite{rodriguez:2008} at the NIST Center for Neutron Research. The neutron momentum transfer is indexed using the Miller indices of  the cubic unit cell, $(h,k,\ell)\!=\!(2\pi/a,2\pi/a,2\pi/a)$, where  $a\!=\!9.66$~\AA{} and all inelastic neutron scattering data has been corrected for the energy dependent neutron absorption from Ir~\cite{Supplemental}.

On CNCS, we used an incident neutron energy of $E_i=3.32$~meV in the high-flux configuration with a chopper frequency of 180~Hz to give an elastic line energy resolution of 0.17~meV (FWHM). A He cryostat and He$^3$ insert was used to cool the sample to a base temperature of $T=0.3$~K. Data were collected for a 360$^\circ$ rotation of the sample with 1 degree step size. Measured neutron count rates were placed into absolute units of the neutron scattering cross-section using incoherent elastic scattering from the sample. The scale factor for conversion to absolute units was additionally cross-checked against the integrated intensity for (200) and (220) nuclear Bragg reflections. All data reduction and analysis was carried out using the Mantid software suite~\cite{Arnold:2014}.

Measurements on MACS were conducted with the sample oriented in the $(h,h,\ell)$ scattering plane using a double-focusing configuration and fixed final energy of 3.7~meV with a BeO filter after the sample and no incident beam filter. The data was corrected for contamination from high-order harmonics in the incident beam neutron monitor.

Unpolarized neutron diffraction measurements were carried out on the HB-1A diffractometer of the High Flux Isotope Reactor(HFIR) at ORNL with collimations of 40’-40’-40’-80’ and fixed incident energy $E_i=14.5$~meV. Polarized neutron diffraction measurements were carried out on the HB-1 triple axis spectrometer of HFIR at ORNL with collimations of 48’-80’-60’-open and fixed incident energy $E_i=13.5$~meV. Both experiments on HB-1A and HB-1 were conducted on the same single crystal with a mass of 42.5~mg aligned in the $(h,k,0)$ scattering plane. 

\tocless{\subsection{X-ray Scattering}}
Resonant elastic x-ray scattering measurements at the Ir L3 edge, $E_{i} = 11.215$~keV, were performed on Beamline 6-ID-B at the Advanced Photon Source. A Pilatus 100K detector with 487$\times$195 pixels of 175$\times$175 micron size, was utilized for the measurement. Polarization analysis was performed using the $(0 0 8)$ reflection of a pyrolytic graphite analyzer and  the Cyberstar scintillator detector. A Joule-Thompson displex cryostat was used to reach a base temperature of 2~K.

\tocless{\subsection{Heisenberg-Kitaev model on the FCC lattice}}
We model the $j=1/2$ doublets of the Ir$^{4+}$ atoms as effective $S=1/2$ spins, $\vec{S}_i$, on an \ac{FCC} lattice with an isotropic $g$-factor. We consider a model with nearest neighbor Heisenberg exchange and Kitaev exchanges~\cite{judd:1959,cook:2015,aczel:2016} 
\begin{equation}
\sum_{\avg{ij}_{\gamma}}\left[
    J\vec{S}_i\cdot \vec{S}_j+
    K S^{\gamma}_i S^{\gamma}_j
    \right] 
\end{equation}
where we have divided the bonds of the \ac{FCC} lattice into three types: $x$, $y$ and $z$, depending on whether they lie in the $yz$, $zx$ or $xy$ planes. When \ac{NNN} interactions are included we consider only a Heisenberg exchange $J_2 \sum_{\avg{\avg{ij}}}\vec{S}_i\cdot\vec{S}_j$. Our best fit parameters (as described in the main text) correspond to $J=0.74$~meV and $K=0.15$~meV.

\tocless{\subsection{Self-consistent non-linear spin-wave theory}}
We consider a semi-classical expansion about the type III ordered state, starting from the $S \rightarrow \infty$ limit. 
We use the Holstein-Primakoff representation~\cite{holstein:1940} of the spin
\begin{align}
  \vec{S}_{i} \equiv &
  \sqrt{S}\left[
    \left(1-\frac{n_{\vec{r}\alpha}}{2S}\right)^{1/2} \nh{a}_{\vec{r}\alpha}  \vhat{e}_{\alpha,-}+
    \h{a}_{\vec{r}\alpha}  \left(1-\frac{n_{\vec{r}\alpha}}{2S}\right)^{1/2}
    \vhat{e}_{\alpha,+}\right] \nonumber \\
  &+\left(S-n_{\vec{r}\alpha}\right)\vhat{e}_{\alpha,0},
\end{align}
where $n_{\vec{r}\alpha} \equiv \h{a}_{\vec{r}\alpha} \nh{a}_{\vec{r}\alpha}$ and $i \equiv \vec{r},\alpha$ labels the unit cell and sublattice of a spin. 
For the collinear type-III orders a four-sublattice unit cell is sufficient for each of the domains. 
The vectors $\vhat{e}_{\alpha,\pm} \equiv (\vhat{x}_\alpha \pm i\vhat{y}_\alpha)/\sqrt{2}$ and $\vhat{e}_{\alpha,0} \equiv \vhat{z}_\alpha$ define a local frame with $\vhat{z}_{\alpha}$ being along the ordering directions of the type III state. 
Expanding in powers of $1/S$ then yields a semi-classical expansion. Including terms up to $O(1)$ we find
\begin{equation*}
  H = N S(S+1) \epsilon_{\rm cl} + H_2  + H_4,
\end{equation*}
where the $O(S^2)$ part, $\epsilon_{\rm cl}$, is the classical ground state energy density and we define the $O(S)$ and $O(1)$ parts in symmetrized form as
\begin{subalign}
  {H}_2 = \frac{1}{2}\sum_{\alpha\beta}\sum_{\vec{k}} & \left[
    {A}_{\vec{k}}^{\alpha\beta} \h{a}_{\vec{k}\alpha}\nh{a}_{\vec{k}\beta} +
    {A}_{-\vec{k}}^{\beta\alpha} \nh{a}_{-\vec{k}\alpha}\h{a}_{-\vec{k}\beta} +\right. \\
    &\left.
           \left(
          {B}^{\alpha\beta}_{\vec{k}}\h{a}_{\vec{k}\alpha}\h{a}_{-\vec{k}\beta} +
          \cb{B}^{\alpha\beta}_{\vec{k}}\nh{a}_{-\vec{k}\beta}\nh{a}_{\vec{k}\alpha}\right)
          \right],\nonumber \\
  {H}_4 = \frac{1}{N_c}\sum_{\alpha\beta\mu\nu} \sum_{\vec{k}\vec{k}'\vec{q}} &\left[
                \frac{1}{(2!)^2} {V}_{\vec{k}\vec{k}[\vec{q}]}^{\alpha\beta\mu\nu} \h{a}_{\vec{k}+\vec{q},\alpha}\h{a}_{\vec{k}'-\vec{q},\beta} \nh{a}_{\vec{k}'\mu} \nh{a}_{\vec{k}\nu} + \right. \\
                & \left. \frac{1}{3!} \left(
                {D}_{\vec{k}\vec{k}'\vec{q}}^{\alpha\beta\mu\nu}\h{a}_{\vec{k}\alpha}\h{a}_{\vec{k}'\beta} \h{a}_{\vec{q}\mu} \nh{a}_{\vec{k}+\vec{k}'+\vec{q},\nu} +\hc
                \right)
                \right].\nonumber
\end{subalign}
In terms of the exchange matrices expressed in these local frames, $\mathcal{J}^{\mu\mu'}_{\vec{\delta},\alpha\alpha'} \equiv \trp{\vhat{e}}_{\alpha,\mu} \mat{J}_{\vec{\delta},\alpha\alpha'}\nh{\vhat{e}}_{\alpha',\mu'}$, the coefficients are
\begin{subalign}
  {A}_{\vec{k}}^{\alpha\beta}
  &=
    S\left(\mathcal{J}^{+-}_{\vec{k},\alpha\beta} - \delta_{\alpha\beta}\sum_\mu \mathcal{J}^{00}_{\vec{0},\alpha\mu}\right), \\
  {B}_{\vec{k}}^{\alpha\beta}
  &=
    S\mathcal{J}^{++}_{\vec{k},\alpha\beta},  \\
  {V}^{\alpha\beta\mu\nu}_{\vec{k}\vec{k}'[\vec{q}]}
  &=
    \left(\delta_{\alpha\mu}\delta_{\beta\nu}\mathcal{J}^{00}_{\vec{k}-\vec{k}'+\vec{q},\alpha\beta}    
    \delta_{\alpha\nu}\delta_{\beta\mu}\mathcal{J}^{00}_{\vec{q},\alpha\beta}\right) \\
    &-
    \left(
    \delta_{\mu\nu}\delta_{\mu\beta}\mathcal{J}^{+-}_{\vec{k}+\vec{q},\alpha\nu}+
    \delta_{\alpha\beta}\delta_{\alpha\mu}\mathcal{J}^{+-}_{\vec{k},\alpha\nu}
    \right), \nonumber \\
  {D}^{\alpha\beta\mu\nu}_{\vec{k}\vec{k}'\vec{q}}
  &=
   -\frac{3}{4} \left(
    \delta_{\alpha\mu}\delta_{\alpha\nu} \mathcal{J}^{++}_{\vec{k}',\beta\alpha}+
    \delta_{\mu\beta}\delta_{\nu\beta} \mathcal{J}^{++}_{\vec{k},\alpha\beta}\right),
\end{subalign}
where the interaction vertices have been left unsymmetrized for brevity. 

At leading order in perturbation theory, these magnon interaction terms renormalize the \ac{LSWT} spectrum, giving corrections to $\delta \mat{A}_{\vec{k}}$ and $\delta \mat{B}_{\vec{k}}$ to $\mat{A}_{\vec{k}}$ and $\mat{B}_{\vec{k}}$ . We treat these corrections self-consistently by writing 
\newcommand{\scavg}[1]{\textrm{MF}}
\begin{subalign}
\delta A^{\alpha\beta}_{\vec{k}} = \frac{1}{N_c} \sum_{\vec{q}\mu\nu}&\Big[
  V^{\alpha \mu \nu \beta}_{\vec{k}\vec{q}[\vec{0}]}\avg{\h{a}_{\vec{q}\mu}\nh{a}_{\vec{q}\nu}}_{\scavg{}} + \nonumber \\ & 
  \frac{1}{2}\left(
  D^{\alpha \mu \nu \beta}_{\vec{k},-\vec{q},\vec{q}}\avg{\h{a}_{-\vec{q}\mu}\h{a}_{\vec{q}\nu}}_{\scavg{}}+
  \cb{D}^{\beta\mu\nu\alpha}_{\vec{k},\vec{q},-\vec{q}}\avg{\nh{a}_{\vec{q}\mu}\nh{a}_{-\vec{q}\nu}}_{\scavg{}}
  \right)
  \Big],\nonumber \\
  \delta B^{\alpha\beta}_{\vec{k}} =
\frac{1}{N_c} \sum_{\vec{q}\mu\nu}&\left[
  D^{\mu\alpha\beta\nu}_{\vec{q},\vec{k},-\vec{k}} \avg{\h{a}_{\vec{q}\mu}\nh{a}_{\vec{q}\nu}}_{\scavg{}} +
  \frac{1}{2}
    V^{\alpha\beta \nu \mu }_{\vec{q},-\vec{q},[\vec{k}-\vec{q}]}\avg{\nh{a}_{\vec{q}\mu}\nh{a}_{-\vec{q}\nu}}_{\scavg{}}
  \right]. \nonumber
\end{subalign}
where $N_c$ is the number of unit cells and the averages $\avg{\cdots}_{\scavg{}}$ are with respect to the quadratic Hamiltonian, $H_2+\delta H_2$, where we have replaced $\mat{A}_{\vec{k}} \rightarrow \mat{A}_{\vec{k}}+\delta \mat{A}_{\vec{k}}$ and $\mat{B}_{\vec{k}} \rightarrow \mat{B}_{\vec{k}}+\delta \mat{B}_{\vec{k}}$. These equations are solved self-consistently (via iteration) to obtain an (effective) corrected quadratic Hamiltonian.

\tocless{\subsection{Calculation of inelastic neutron scattering intensity}}
The inelastic neutron scattering intensity is computed theoretically in terms of the spin-spin-correlation function
\begin{equation}
\mathcal{S}_{\mu\nu}(\vec{k},\omega) \equiv
\frac{1}{2\pi N} \int dt e^{-i\omega t}
\avg{M^\mu_{-\vec{k}} M^\nu_{\vec{k}}(t)}
\end{equation}
where $\vec{M}_{\vec{k}} \equiv g\mu_B \sum_i e^{-i\vec{k}\cdot \vec{r}_i} \vec{S}_i $ is the magnetization
operator at wave-vector $\vec{k}$ and $N$ is the total number of spins. The observed intensity is given by
\begin{align}
    I(\vec{k},\omega) &\propto F(k)^2 \sum_{\mu\nu} \left(\delta_{\mu\nu}-\hat{k}_{\mu}\hat{k}_{\nu}\right) \mathcal{S}_{\mu\nu}(\vec{k},\omega)
\end{align}
where $F(k)$ is the Ir$^{4+}$  magnetic form factor. 

Within linear spin-wave theory the dynamical structure factor can be expressed in terms of the transverse-transverse part of the spin-spin correlation function. This remains true in our non-linear spin-wave theory (up to an overall intensity renormalization) and $I(\vec{k},\omega)$ can be computed using the self-consistently determined $H_2+\delta H_2$ as in \ac{LSWT}. For more details, including the averaging needed to emulate the binning of the experimental data, we refer the reader to the Supplemental Material~\cite{Supplemental}.

\tocless{\subsection{Classical spin dynamics simulations}}
In the paramagnetic phase we consider the \ac{FCC} \ac{HK} model with $J=0.74$~meV and $K=0.15$~meV in the classical limit where the spins are unit length vectors $|\vec{S}_i|^2=1$. 
The spins obey conventional Landau-Lifshitz dynamics
\begin{equation}
    \label{eq:lldyn}
    \frac{d\vec{S}_i}{dt} = -\vec{S}_i \times \frac{\del H}{\del \vec{S}_i},
\end{equation}
where $-\del H/\del \vec{S}_i \equiv \vec{B}_i$ is the (local) exchange field due to the neighboring spins. The initial conditions $\vec{S}_i(0)$ are drawn from a thermal distribution at temperature $T$ using Monte Carlo sampling. Once a sample of trajectories $\vec{S}_i(t)$ are obtained, the part of the dynamical structure factor relevant for inelastic neutron scattering is given by
\begin{equation}
\mathcal{S}_{\rm cl}(\vec{k},\omega) = \sum_{\mu\nu} \left(\delta_{\mu\nu} - \hat{k}_\mu\hat{k}_\nu \right) \avg{\bar{S}^\mu_{\vec{k}}(\omega) S^{\nu}_{\vec{k}}(\omega)},
\end{equation}
where $\vec{S}_{\vec{k}}(\omega)$ is the Fourier transform of $\vec{S}_i(t)$ in both space and time. 
To account for \KIC{} being $S=1/2$ we rescale frequencies by a factor of $S$, ensuring that in the low temperature limit the classical and quantum spin-wave frequencies agree. To partially account for quantum effects we also multiply by an energy dependent correction factor~\cite{zhang:2019, scheie:2022}
\begin{equation}
F_{\rm qu}(\omega) \equiv \beta\omega\left(1+n_B(\omega)\right),
\end{equation}
where $n_B(\omega) = 1/(e^{\beta\omega}-1)$ is the Bose distribution.

Thermal samples are generated using Monte Carlo with single-site heat-bath~\cite{miyatake:1986} and over-relaxation~\cite{creutz:1987} updates, annealing down from high temperature $O(10J)$ to the temperature of interest and then thermalizing for the same number of sweeps. For Fig.~\ref{fig:para} in the paramagnetic phase only a small number of sweeps, typically $O(10^3)$, are necessary to reach equilibrium and the system size used was $N=L^3$ with $L=32$. For each initial state we solve the coupled non-linear ordinary differential equations using an adaptive fourth-order Runge-Kutta method. At the temperatures of interest $O(10^2)$ samples are sufficient to reach convergence in both energy resolved and energy averaged quantities. The adaptive time stepping was performed with (absolute and relative) error tolerances of $10^{-8}$. 

Simulations used to establish the dependence of the ordering type and ordering temperature on $J_2/J$ used a larger number of sweeps, $O(10^5)$, on smaller system sizes with a conventional cell $N=4L^3$ where $L=8$, $10$, $12$, as well as parallel tempering to aid in reaching equilibrium. The phase transition temperature was inferred from the location of a sharp maximum in the heat capacity and the onset of a type III order parameter.

\tocless{\putbib}

\end{bibunit}

\cleardoublepage
\onecolumngrid

\setcounter{section}{0}
\setcounter{table}{0}
\setcounter{figure}{0}
\setcounter{equation}{0}
\setcounter{page}{1}
\renewcommand{\theequation}{S\arabic{equation}}
\renewcommand{\thefigure}{S\arabic{figure}}
\renewcommand{\thetable}{S\arabic{table}}

\begin{bibunit}
\begin{large}
    \begin{center}
\textbf{Supplemental Material for ``Pulling order back from the brink of disorder: Observation of a nodal line spin liquid and fluctuation stabilized order in \KIC{} ''}
\end{center}
\end{large}
\vspace{0.5cm}
\tableofcontents
\section{Experiment}
\subsection{Sample characterization}
Powder X-ray diffraction measurements were carried out on the 11-BM diffractometer at the Advanced Photon Source using   an X-ray wavelength of 0.45893\textup{~\AA}. Measurements were conducted at at $T=90$~K and 300~K. Reitveld refinement of powder data was carried out using the FullProf software~\cite{rodriguez:1993}. 

\begin{figure}[h!]
    \centering
    \includegraphics[width = 0.65\textwidth]{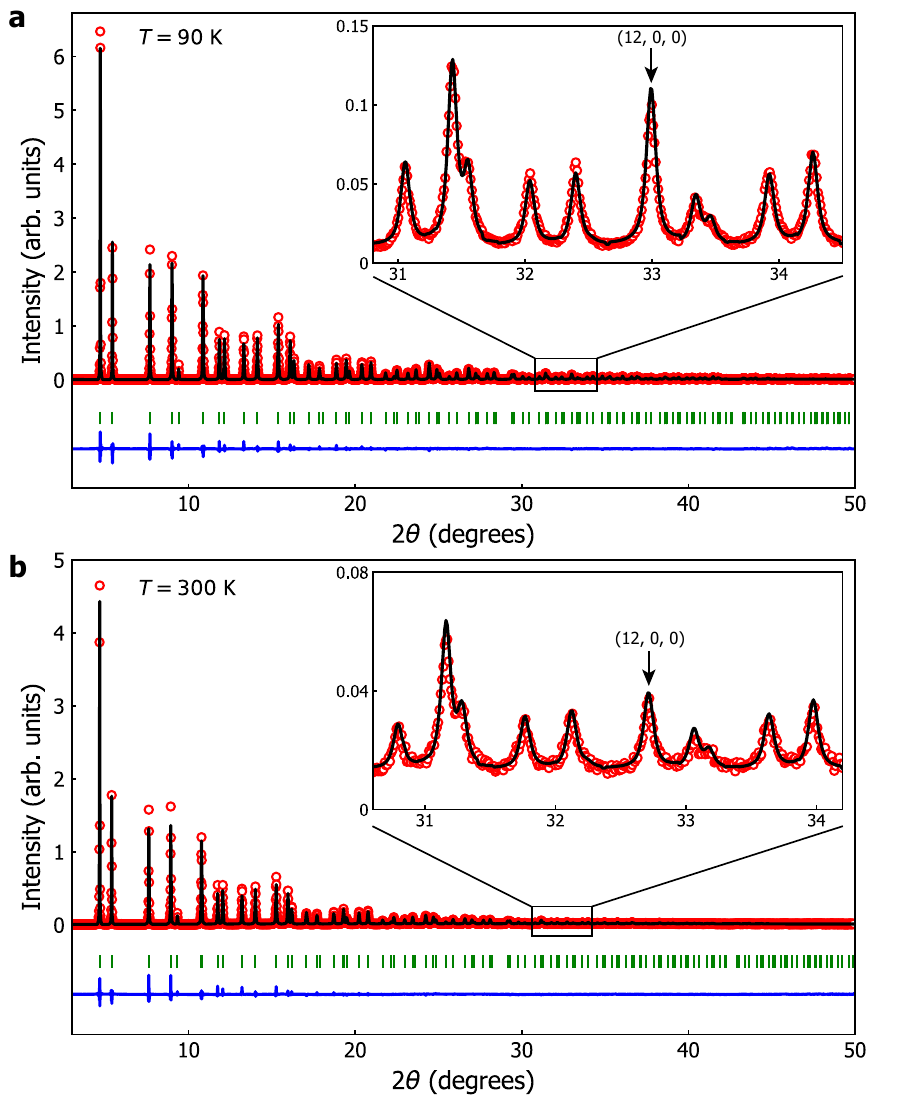}
    \caption{\label{fig:SM1}{ High-resolution synchrotron powder X-ray diffraction on \KIC{}.} Results of a Rietveld refinement(black line) are displayed overtop of the high-resolution powder x-ray diffraction (red circle)  at ({a}) 90~K and ({b}) 300~K . The expected reflections are indexed with green ticks and the blue line is the fit residual. Insets show the zoom-in of the high-$Q$ diffraction pattern. No structural distortion or impurity phases are discernible.}
\end{figure}

Fig.~\ref{fig:SM1} shows the diffraction data and Rietveld refinement at 90~K (\textbf{a}) and 300~K (\textbf{b}).  \KIC{} maintains a cubic crystal structure at all measurement temperatures with no observable symmetry lowering  down to 90~K. Table~\ref{tab:table1} summarizes the resulting crystallographic parameters and atomic positions. All refined parameters are consistent with  published reports~\cite{reig-i-plessis:2020, khan:2019}.

\begin{table}[h!]
\caption{\label{tab:table1}%
Crystallographic parameters and details of the structure refinement for \KIC{}. }
\begin{ruledtabular}
\begin{tabular}{l l l}
\textrm{$T$}&
\textrm{90 K}&
\textrm{300 K}\\
\colrule
Space group & $Fm\overline{3}m$ & $Fm\overline{3}m$\\
$a=b=c$ (\AA) & 9.6973(2) & 9.7777(3)\\
$\alpha=\beta=\gamma$ ($^{\circ}$) & 90\degree & 90\degree\\
\colrule
Atomic parameters\\
\colrule
K & $x/a=0.25, y/b=0.25, z/c=0.25$ & $x/a=0.25, y/b=0.25, z/c=0.25$\\
 & $B_{\rm iso}=1.119(39)$ & $B_{\rm iso}=2.779(68)$\\
 & ${\rm Occ}=1.003$ & ${\rm Occ}=0.994$\\
\colrule
Ir & $x/a=0, y/b=0, z/c=0$ & $x/a=0, y/b=0, z/c=0$\\
 & $B_{\rm iso}=0.260(8)$ & $B_{\rm iso}=0.886(10)$\\
 & ${\rm Occ}=1.002$ & ${\rm Occ}=1.001$\\
\colrule
Cl & $x/a=0.2394(2), y/b=0, z/c=0$ & $x/a=0.2373(2), y/b=0, z/c=0$\\
 & $B_{\rm iso}=0.840(28)$ & $B_{\rm iso}=2.203(47)$\\
 & ${\rm Occ}=0.991$ & ${\rm Occ}=0.998$\\
 \colrule
 Refinement\\
 \colrule
 $R_{\rm p}$ & 9.32 & 8.59\\
 $R_{\rm exp}$ & 4.90 & 5.57\\
 $\chi^{2}$ & 6.72 & 3.52\\

\end{tabular}
\end{ruledtabular}
\end{table}

\begin{figure}[h!]
    \centering
    \includegraphics[]{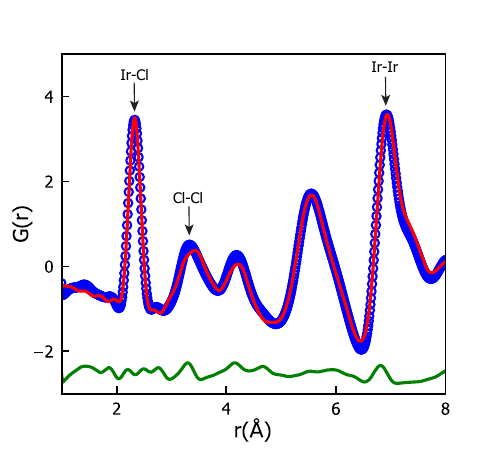}
    \caption{\label{fig:SM2}{ X-ray pair distribution function (PDF) for \KIC{} measured at 300~K.} The measured PDF (red circles) on top of the refinement (blue line) with the difference (green line). The positions for distances of Ir-Cl, Cl-Cl and Ir-Ir are pointed out. No deviation from the cubic structure is observed at room temperature.}
\end{figure}

To complement the diffraction measurements and verify a local cubic structure in \KIC{}, $T=300$~K x-ray pair distribution function (PDF) measurements were carried out at the Advanced Photon Source, 11-ID-B with an X-ray wavelength of 0.2116\textup{~\AA}. We used GSAS-II~\cite{Toby:2013} and PDFgui~\cite{farrow:2007} for data processing and refinement. The measured PDF and refinement are shown in Fig.~\ref{fig:SM2} with positions for distances of Ir-Cl, Cl-Cl and Ir-Ir indicated out. The data reveal a single nearest neighbor Ir-Cl distance as expected for a local cubic environment. No deviation from the cubic structure is observed at room temperature. 

Heat capacity measurements were conducted using a Quantum Design PPMS in zero field during a warming process. The phonon contribution was estimated using the empirical model by N. Khan \textit{et al}~\cite{khan:2019}, which incorporates one Debye-type and three Einstein-type terms. The Debye term accounts for 24\% of the total modes, characterized by a Debye temperature $\Theta_{\rm D}=91.1$~K. The three Einstein terms have Einstein temperatures $\Theta_{\rm E}=459,~242,~136$~K, respectively.

\begin{figure}[h!]
    \centering
    \includegraphics[]{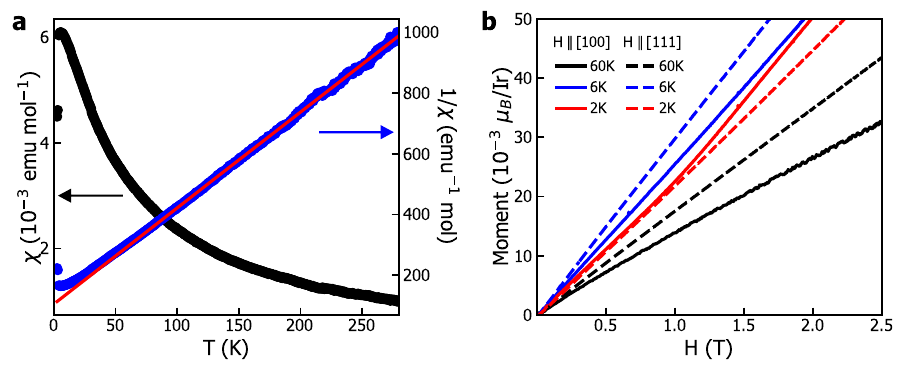}
    \caption{\label{fig:SM3}{ Magnetic properties of \KIC{}.} {(a)}, Temperature dependence of magnetic susceptibility (black curve) and inverse magnetic susceptibility (blue curve) for \KIC{} measured under a magnetic field of $H=0.1$~T. The red line is the Curie-Weiss fit to the inverse susceptibility data at $T>100$~K. {(b)}, Field dependence of the magnetic moment at 2~K (red), 6~K (blue) and 60~K (black) along $[1,0,0]$ (solid) and $[1,1,1]$ (dashed) directions of the crystal lattice.}
\end{figure}

Magnetization measurements were carried out using a Quantum Design PPMS vibrating sample magnetometer (VSM). Fig.~\subref{fig:SM3}{(a)} shows the temperature dependent magnetic susceptibility measured under a $[1,0,0]$ oriented $H=0.1$~T field. Data was collected on warming after zero-field cooling. A Curie-Weiss temperature of $\theta_{\rm CW} = -32.98$~K and paramagnetic moment of 1.59~$\mu_B$ were extracted from a Curie-Weiss fit to the data  over the temperature range $100 < T <280$~K. These parameters are in agreement with published characterization~\cite{khan:2019}. Fig.~\subref{fig:SM3}{(b)} shows the field dependent magnetization at 2~K, 6~K, 60~K with the field along $[1,0,0]$ and $[1,1,1]$ crystal  directions.

\subsection{Details of Absorption Corrections for Inelastic Neutron Scattering Data}
Iridium has a large neutron absorption cross-section of 425 barn.  The absorption lengths of \KIC{} are 4.2 and 8.2~cm$^{-1}$ at neutron energies 14.5 and 3.32~meV, respectively. Thus an absorption correction is necessary in processing the neutron scattering data. The absorption correction for CNCS data was conducted using the built-in Mantid algorithm~\cite{arnold:2014}. The Mantid algorithm relies on a numerical integration method, which effectively calculates the attenuation factors arising from sample absorption based on its given material properties. In this study, an approximate cylindrical crystal array with a radius of 0.12 cm and a height of 3 cm was utilized for the measurements. The cylindrical crystal array was discretized into small cubes, each measuring $1\times 1 \times 1$~mm in size. The cubes whose centers lie within the sample make up the set of integration elements. Path lengths through the sample were then computed for each center-point of the selected integration elements. Finally, a numerical integration technique was employed over the volume elements using these calculated path lengths to determine the necessary corrections~\cite{arnold:2014}.

\subsection{Additional Inelastic Neutron Scattering Data}
Additional inelastic neutron scattering measurements carried out on the MACS spectrometer with the \KIC{} crystal aligned in the $[h,h,l]$ scattering plane are shown in Fig.~\ref{fig:SM2}. The observed spin wave dispersion and gap are consistent with the CNCS data and non-linear spin wave modeling of the nearest neighbor Heisenberg-Kitaev model described below and in the main text.
\begin{figure}[h!]
    \centering
    \includegraphics[]{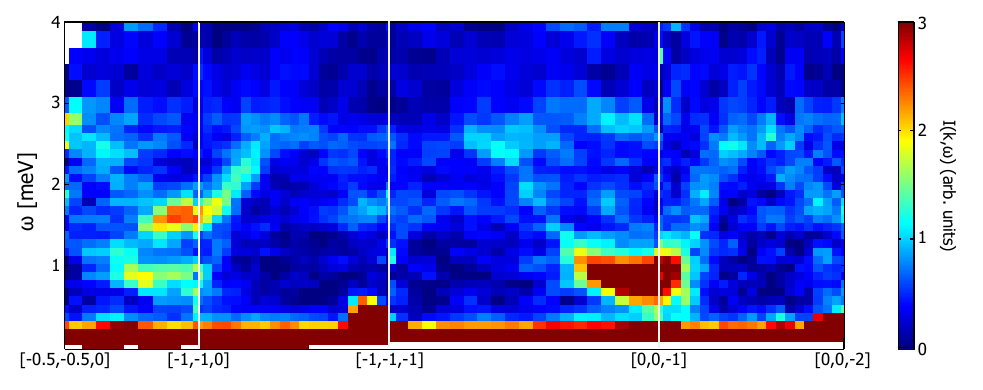}
    \caption{\label{fig:SM4}{Additional inelastic neutron scattering data.} Magnetic excitation spectra measured in the $[h,h,l]$ scattering plane on MACS at 1.8 K.} 
\end{figure}

\subsection{Additional Polarized Neutron Scattering Data}
\begin{figure}[h!]
    \centering
    \includegraphics[]{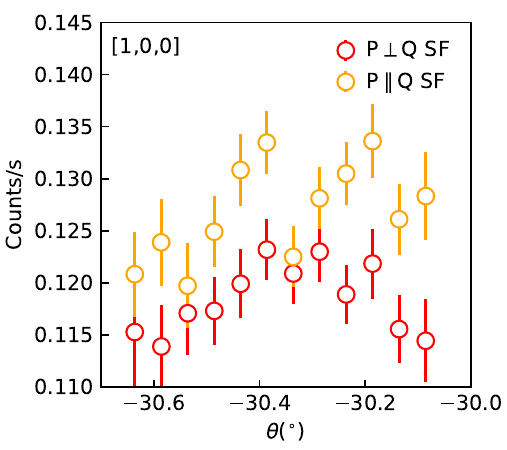}
    \caption{\label{fig:SM5}{Polarized neutron diffraction measurement on the type-I peak at 1.5~K.} The spin-flip channel was measured for neutron spins parallel to the scattering vector ($\vec{P} \parallel \vec{Q}$) and perpendicular to the scattering plane ($\vec{P} \perp \vec{Q}$). The difference between the two channels corresponds to the out-of-plane component.}
\end{figure}

 Polarized neutron scattering measurement on the type-I magnetic peak [1,0,0] are shown in Fig.~\ref{fig:SM5}. The spin-flip intensity of [1,0,0] reflection is sensitive to both in and out-of-plane components, i.e. [0,0,1] direction of the magnetization for guide fields $\vec{P} \parallel \vec{Q}$, and only in-plane components for $\vec{P}\perp \vec{Q}$. Although the [1,0,0] peak is extremely weak, we still observe a stronger intensity in the $\vec{P} \parallel \vec{Q}$ channel compared with the $\vec{P}\perp \vec{Q}$ channel. Based on a finite flipping ratio of 15.2 which gives rise to the intensity in the $\vec{P}\perp \vec{Q}$ channel, a total background of 0.126 counts/s is estimated in the $\vec{P} \parallel \vec{Q}$ channel, and thus an intensity of $\sim$ 0.008 counts/s is left for the out-of-plane component of magnetization from type-I magnetic order.

\subsection{Survey of order by disorder gaps in various materials}

In order to place our observations and the scale of quantum fluctuations in \KIC{} in the broader context of order by quantum disorder candidate materials,  below we summarize magnon bandwidths and excitation gaps that have been attributed to quantum order by disorder. 

\begin{table}[h!]
\caption{\label{tab:table2}%
Scale of quantum order by disorder gap relative to bandwidth in candidate materials.
}
\begin{ruledtabular}
\begin{tabular}{c c c c c c}
\textrm{Material}&
\textrm{Spin}&
\textrm{$W$ (Bandwidth)}&
\textrm{$\Delta$ (Gap size)}&
\textrm{$\Delta/W$}&
\textrm{Ref.}\\
\colrule
Ca$_3$Fe$_2$Ge$_3$O$_{12}$ & $S=5/2$ & $\sim$ 1.95~meV & 136~$\mu$eV & 7.0\% &~\cite{brueckel:1988} \\
CoTiO$_3$ & $S_{\rm eff} = 1/2$ &$\sim$ 12~meV & 1~meV & 8.3\% &~\cite{yuan:2020, elliot:2021}\\
ErTi$_2$O$_7$ & $S_{\rm eff}=1/2$ &$\sim$ 0.45~meV & 43~$\mu$eV & 9.6\% &~\cite{ruff:2008, champion:2003, zhitomirsky:2012, savary:2012} \vspace{0.5em} \\
\hline
\KIC{}& $j=1/2$ & 2.5~meV & 0.7~meV & 30\% & This study
\end{tabular}
\end{ruledtabular}
\end{table} 

\section{Theory}
\subsection{Model}
We consider a minimal model for the (effective) $j=1/2$ doublets~\cite{jackeli:2009} of the Ir\tsup{4+} ions on an FCC lattice relevant for K\tsub{2}IrCl\tsub{6}. The degrees of freedom of these doublets are described by an effective $S=1/2$ spin, $\vec{S}_i$, with an isotropic $g$-factor. From electron spin resonance measurements~\cite{bhaskaran:2021} we expect the $g$-factor to be $g \approx 1.8$, not far from $g=2$ expected for an ideal $j_{\rm eff}=1/2$ doublet~\cite{Note1}.

The cubic symmetry  strongly constrains the symmetry-allowed exchange  interactions.
At nearest-neighbor level, this includes nearest neighbor Heisenberg exchange as well as bond-dependent Kitaev and
$\Gamma$ exchanges~\cite{judd:1959,cook:2015,aczel:2016}
\begin{equation}
\sum_{\avg{ij}_{\alpha\beta(\gamma)}}\left[
    J\vec{S}_i\cdot \vec{S}_j+
    K S^{\gamma}_i S^{\gamma}_j +
    (-1)^{\sigma^{\alpha\beta}_{ij}} \Gamma \left(S^{\alpha}_i S^{\beta}_j + S^{\alpha}_i S^{\beta}_j\right)
    \right] 
    \label{supp:eq:jkg}
\end{equation}
We have divided the bonds of the lattice into three types: $x$, $y$ and $z$,
depending on whether they lie in the $yz$, $zx$ or $xy$ planes.  The
sign of the $\Gamma$ term is determined by the bond direction
$\vec{d}_{ij} \equiv \vec{r}_j -\vec{r}_i$ as
$\sigma^{\alpha\beta}_{ij} \equiv {\rm sgn}(d^\alpha_{ij}
d^\beta_{ij})$. Second neighbor interactions along the cubic axes are restricted to
only a Heisenberg or Kitaev exchange~\cite{diop:2022}
\begin{equation}
J_2 \sum_{\avg{\avg{ij}}}\vec{S}_i\cdot\vec{S}_j
+K_2 \sum_{\avg{\avg{ij}}_{\gamma}}S^{\gamma}_i S^{\gamma}_j
    \label{supp:eq:j2}
\end{equation}
where $\gamma$ is the bond direction. For simplicity, we will restrict our discussion to only isotropic second neighbor exchange setting $K_2=0$. 

\subsection{Classical Ground States}
\subsubsection{Luttinger-Tisza}
\begin{figure}[t]
    \centering
    \includegraphics[width=0.66\textwidth]{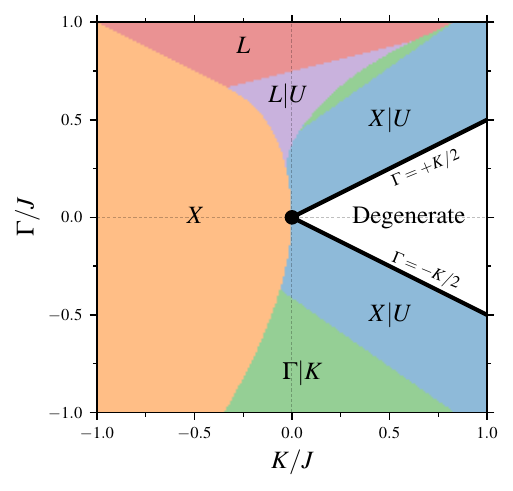}
    \caption{Phase diagram for $J$-$K$-$\Gamma$ model [Eq.~(\ref{supp:eq:jkg})] with $J>0$ from the Luttinger-Tisza approach. For each phase we have indicated the minimal wave-vector, with $X \equiv [1,0,0]$, $L \equiv [0.5, 0.5, 0.5]$, $K \equiv [\tfrac{3}{2},\tfrac{3}{2},0]$, $\Gamma \equiv [0,0,0]$ and $U \equiv [1, 0.25, 0.25]$. The notation $X|U$ denotes a minimum wave-vector along the line connecting $X$ and $U$ (for example). The region with $|\Gamma|<K/2$, $K>0$ does not have a unique minimum, but instead a degenerate line of minima along $X|W$ where $W \equiv [0.5, 1, 0]$. Note that we have distinguished phases only by wave-vector, not by moment direction.}
    \label{supp:fig:luttinger-tisza}
\end{figure}
To understand the qualitative features of the classical phase diagram, we begin by revisiting the Luttinger-Tisza method~\cite{luttinger:1946,litvin:1974} that has been used in previous works~\cite{kimchi:2014,cook:2015,khan:2019,diop:2022}. We find a rich phase diagram including multiple commensurate and incommensurate magnetic phases, as well as a broad region with a degeneracy along one-dimensional wave-vector manifolds -- ``spiral lines'' along $[q,1,0]$, $[0,q,1]$, $[1,0,q]$ and equivalents. This is illustrated in Fig.~\ref{supp:fig:luttinger-tisza}. The commensurate phase with wave-vector $X$ is a type I phase, while the type III phase would correspond to a wave-vector of $W$. We note that the incommensurate phases do not always satisfy the spin-length constraint and thus do not always represent true classical ground states.~\cite{Note2} In these regions direct simulation via parallel tempering Monte Carlo~\cite{hukushima:1996,swendsen:1986,marinari:1992} using heat-bath updates~\cite{miyatake:1986} and iterative minimization~\cite{sklan:2013} yield a complex set of multi-$Q$ incommensurate spirals whose wave-vectors qualitatively track the Luttinger-Tisza result.

\subsubsection{Ground State Manifold}
\label{supp:sec:ground-state-manifold}
\begin{figure}[t]
    \centering
    \includegraphics[width=\textwidth]{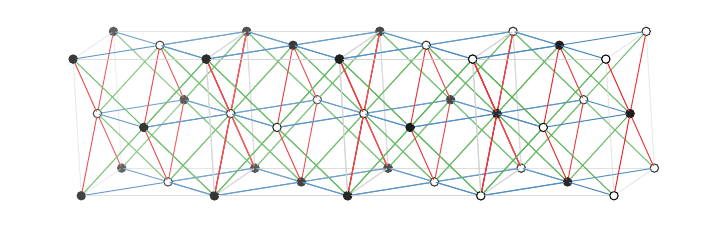}
    \caption{Illustration of a generic ``stacked'' ground state (along $\vhat{x}$) of the nearest-neighbor model [Eq.~(\ref{supp:eq:jkg})] when $J>0$ and $|\Gamma|<K/2$,. The moment direction would be parallel to stacking direction, $\vhat{x}$, with filled and open circles denoting the sign.}
    \label{supp:fig:stacked}
\end{figure}
In the region with $K>0$ and $|\Gamma|<K/2$ containing the ``spiral lines'' the Luttinger-Tisza energy is $E_0=-(2J+K)S^2$, representing a lower-bound on the true classical ground state energy of the system. One can find classical ground states that saturate this bound and thus represent true ground states of the model. We first construct a sub-extensive number of collinear states with this energy. These can be understood as an arbitrary stacking of N\'eel planes along the cubic directions, with two choices for the N\'eel state in each plane. With three choice of stacking direction ($x$, $y$ or $z$) and two choices per plane, we have $3 \cdot 2^{2L}$ states where $L$ is the number of cubic unit cells along the stacking direction. More explicitly we can define three families of states
\begin{align*}
  \vec{S}^x_{i} &= S(-1)^{n_1+n_2} \sigma_{n_2+n_3}^x \vhat{x}, \\
  \vec{S}^y_{i} &= S(-1)^{n_2+n_3} \sigma_{n_3+n_1}^y \vhat{y}, \\
  \vec{S}^z_{i} &= S(-1)^{n_3+n_1} \sigma_{n_1+n_2}^z \vhat{z}
\end{align*}
where we have expressed the position $\vec{r}_i = n_1\vec{a}_1+n_2\vec{a}_2+n_3\vec{a}_3$
in terms of the primitive lattice vectors $\vec{a}_1$, $\vec{a}_2$ and $\vec{a}_3$. The $\sigma^{\mu}_n$ are Ising variables equal to $\pm 1$ that encode the choice N\'eel plane. For each of these
states the inter-plane couplings cancel and only the plane perpendicular to the stacking direction contributes to the classical energy. For $|\Gamma|<K/2$ this is minimized by a N\'eel state (satisfying the Heisenberg part) with moment parallel to the stacking direction (satisfying the Kitaev part), with $\Gamma$ dropping out. This gives the required $E = -(2J+K)S^2$. These ``stacked'' states are identical those that make up the  
sub-extensive discrete degeneracy of the FCC Ising anti-ferromagnet~\cite{luttinger:1950}.

This discrete degeneracy does not exhaust the ground state manifold. As these stacked states are collinear and can be oriented along perpendicular axes, linear combinations of these states are also ground states. More explicitly, if we define
$$
\vec{S}_{i} \equiv S\left[
\alpha_x (-1)^{n_1+n_2} \sigma_{n_2+n_3}^x \vhat{x}+
\alpha_y (-1)^{n_2+n_3} \sigma_{n_3+n_1}^y \vhat{y}+
\alpha_z (-1)^{n_3+n_1} \sigma_{n_1+n_2}^z \vhat{z}\right]
$$
where $\sum_{\mu}\alpha_{\mu}^2=1$, then we can see that these are normalized $|\vec{S}_{i}|^2=1$. The cross terms in the energy vanish for distinct stackings, yielding the same $-(2J+K)S^2$ energy. We have thus established a
very large \emph{continuous} manifold of ground states: for each
discrete ground state (say) stacked along the $\vhat{z}$ direction, we
can ``rotate'' it into two arbitrary ground states stacked in the
$\vhat{x}$ and $\vhat{y}$ directions, producing a new (non-collinear) ground state.

This continuous manifold is a subset of the continuous manifold found in the Heisenberg limit when $J>0$ and $K=\Gamma=0$~\cite{anderson:1950,alexander:1980}. In the Heisenberg limit, this manifold is considerably larger. For example, the stacked N\'eel planes can be mixed with states stacked along the same direction, but with orthogonal moment directions.

\subsubsection{Effect of Second-neighbor Exchange}
\label{supp:sec:effect-of-finite-j2}
\begin{figure}[t]
    \centering
     \centering
  \begin{overpic}[width=0.32\textwidth]{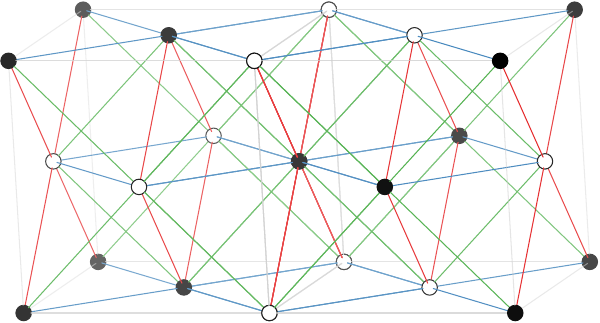}\put(42,-5){$X$}\end{overpic}
  \begin{overpic}[width=0.23\textwidth]{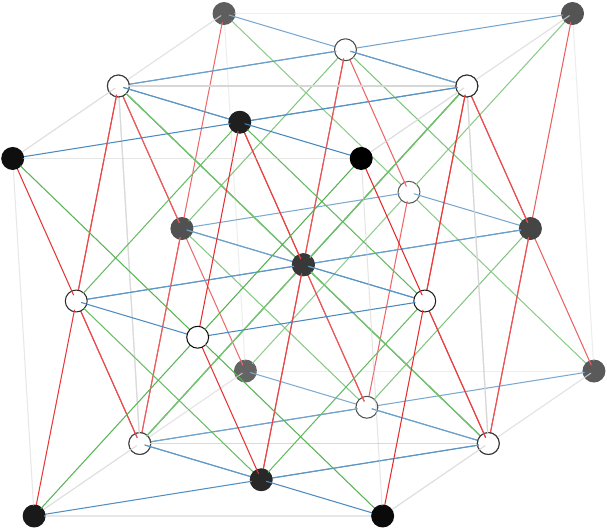}\put(42,-7){$Y$}\end{overpic}
  \begin{overpic}[height=0.32\textwidth]{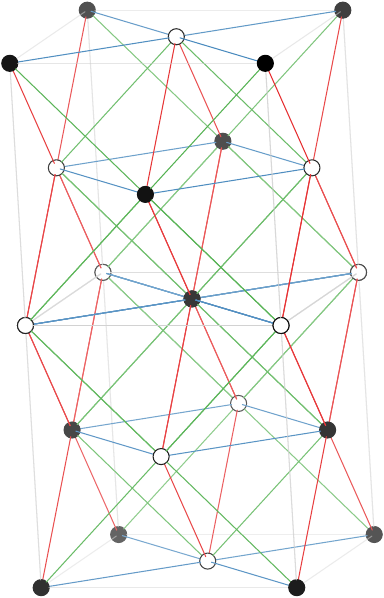}\put(42,-5){$Z$}\end{overpic}
  \vspace{0.5em}
    \caption{Illustration of the sign structure of type-III anti-ferromagnetic states favored by anti-ferromagnetic second-neighbor exchange, $J_2>0$. Filled circles indicate moments along the direction of the stacked N\'eel planes, open circles indicate moments against the stacking direction.}
    \label{supp:fig:type-III}
\end{figure}
\begin{figure}[t]
      \centering
  \begin{overpic}[width=0.2\textwidth]{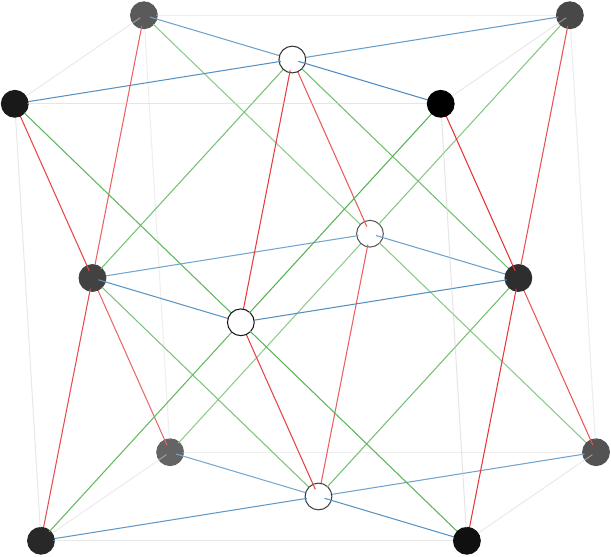}\put(42,-10){$[100]$}\end{overpic}
  \begin{overpic}[width=0.2\textwidth]{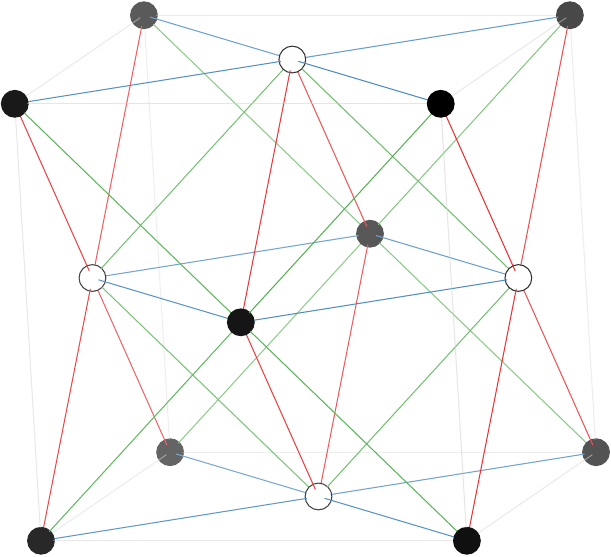}\put(42,-10){$[010]$}\end{overpic}
  \begin{overpic}[width=0.2\textwidth]{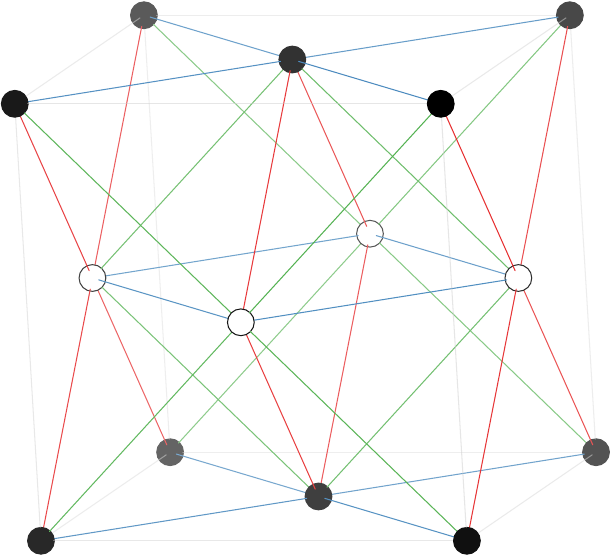}\put(42,-10){$[001]$}\end{overpic}
    \caption{Illustration of the sign structure of the type-I anti-ferromagnetic states favored by ferromagnetic second-neighbor exchange, $J_2<0$. Filled circles indicate moments along the direction of the stacked N\'eel planes, open circles indicate moments against the stacking direction. }
    \label{supp:fig:type-I}
\end{figure}
Adding a finite second neighbor coupling immediately lifts most, but not all, of the large accidental degeneracy present in the nearest-neighbor model with $K>0$ and $|\Gamma|<K/2$. 

Antiferromagnetic second neighbor exchange, $J_2>0$, favors a collinear type-III anti-ferromagnet containing wave-vectors $(0.5,1,0)$, $(0, 0.5, 1)$, $(1,0,0.5)$ or equivalents. This can be viewed as a period four stacking of the form $(\pm,\pm,\mp,\mp)$ or equivalent of N\'eel planes, with the moment direction parallel to the stacking direction. There are four choices for the periodicity and three stacking directions and thus 12 distinct collinear states. The sign structure of this state is illustrated in Fig.~\ref{supp:fig:type-III}.

For $J_2<0$ it favors a collinear type-I anti-ferromagnet with wave-vectors $(1,0,0)$, $(0,1,0)$ or $(0,0,1)$. These can be viewed as stacks of \emph{ferromagnetic} planes with the moments oriented perpendicular to the stacking direction. There are 12 distinct collinear type I states, corresponding to three stacking directions, two choices of moment direction and two choices for the alternation pattern. In terms of stacking of \emph{N\'eel} planes, these can be viewed as being composed of the four stackings of the form $(\pm,\pm)$ or $(\pm,\mp)$ along each of the three cubic directions with the moment parallel to the stacking direction. The sign structure of this state is illustrated in Fig.~\ref{supp:fig:type-I}.

In both cases there is a remnant of the larger degeneracy of the nearest-neighbor model: states stacked along different directions can be continuously mixed. For $J_2>0$ one can still mix a type-III state with any other type-III state with perpendicular moment, and for $J_2<0$ you can smoothly mix a type-I state with another type-I state so along as their moments are orthogonal. This degeneracy cannot be lifted by any exchange interactions bilinear in the spins that respect the cubic symmetry of the crystal. For each case, three orthogonal type I or type III states can be grouped into a order parameter $\vec{m}$ that transforms as a vector under the cubic symmetries. Since the only bilinear in $\vec{m}$ that respects cubic symmetry is $\propto |\vec{m}|^2$ we see that an accidental $O(3)$ will remain regardless of the addition of further (symmetry preserving) exchange interactions~\cite{savary:2012}.

\subsection{Linear Spin Wave Theory}
\label{supp:sec:lswt}
We now consider a semi-classical expansion about the one of ground states described above. The spin operators can expressed in terms of Holstein-Primakoff bosons as
\begin{equation}
  \vec{S}_{\vec{r}\alpha} \equiv
  \sqrt{S}\left[
    \left(1-\frac{n_{\vec{r}\alpha}}{2S}\right)^{1/2} \nh{a}_{\vec{r}\alpha}  \vhat{e}_{\alpha,-}+
    \h{a}_{\vec{r}\alpha}  \left(1-\frac{n_{\vec{r}\alpha}}{2S}\right)^{1/2}
    \vhat{e}_{\alpha,+}\right]
  +\left(S-n_{\vec{r}\alpha}\right)\vhat{e}_{\alpha,0},
\end{equation}
where $n_{\vec{r}\alpha} \equiv \h{a}_{\vec{r}\alpha} \nh{a}_{\vec{r}\alpha}$ and $\vec{r},\alpha$ denotes the unit cell and sublattice index of the spin. For the collinear type-I or type-III orders a four-sublattice unit cell is sufficient, while for other stacked states or for non-collinear mixtures of type-I and type-III larger unit cells are necessary.
The vectors
$\vhat{e}_{\alpha,\pm}$, $\vhat{e}_{\alpha,0}$ define a local frame of reference; in a more conventional
Cartesian basis one defines $\vhat{e}_{\alpha,\pm} \equiv (\vhat{x}_\alpha \pm i\vhat{y}_\alpha)/\sqrt{2}$ and
$\vhat{e}_{\alpha,0} \equiv \vhat{z}_\alpha$. 
It is useful to write the exchange matrix in the frame aligned with these axes 
\begin{equation}
  \label{eq:lswt:exchange}
\mathcal{J}^{\mu\mu'}_{\vec{\delta},\alpha\alpha'} \equiv \trp{\vhat{e}}_{\alpha,\mu} \mat{J}_{\vec{\delta},\alpha\alpha'}\nh{\vhat{e}}_{\alpha',\mu'}.
\end{equation}
Expanding in powers of $1/S$ then yields a semi-classical
expansion about the ordered state defined by $\vhat{e}_{\alpha,0}$,
typically chosen to be along the classical ordering direction.

At order $O(S)$ in the Holstein-Primakoff operators we have
\begin{equation}
  \vec{S}_{\vec{r}\alpha} \approx
  \sqrt{S}\left[
   \nh{a}_{\vec{r}\alpha}  \vhat{e}_{\alpha,-}+
    \h{a}_{\vec{r}\alpha} 
    \vhat{e}_{\alpha,+}\right]
  +\left(S-n_{\vec{r}\alpha}\right)\vhat{e}_{\alpha,0},
\end{equation}
Inserting this into our spin Hamiltonian and keeping only terms
to $O(S)$ yields
\begin{equation}
  \label{eq:lswt:ham}
  H = N S(S+1)\epsilon_{\rm cl} +\frac{1}{2}  \sum_{\vec{k}}
  \left(\trp{[\h{\vec{a}}_{\vec{k}}]}\ \trp{\vec{a}}_{-\vec{k}}\right)
  \left(
    \begin{array}{cc}
      \mat{A}_{\vec{k}} & \mat{B}_{\vec{k}} \\
      \cb{\mat{B}}_{-\vec{k}} & \cb{\mat{A}}_{-\vec{k}}
    \end{array}
  \right)
  \left(
  \begin{array}{c}
    \nh{\vec{a}}_{\vec{k}} \\
    \h{\vec{a}}_{-\vec{k}}
  \end{array}\right) + O(S^{1/2}),
\end{equation}
where $N_s$ is the number of sublattices, $N$ is the total number of
sites and we have defined the Fourier transforms of the bosons as
$a_{\vec{k}\alpha} \equiv N_c^{-1/2} \sum_{\vec{r}}
e^{-i\vec{k}\cdot\vec{r}} a_{\vec{r}\alpha}$ where $N = N_c N_s$. 
The classical energy per site is defined as
\begin{equation}
  \label{eq:lswt:classical}
  \epsilon_{\rm cl} \equiv \frac{1}{2N_s}\sum_{\alpha\alpha'} \sum_{\vec{\delta}} \mathcal{J}^{00}_{\vec{\delta},\alpha\alpha'}, 
\end{equation}
and the matrices $\mat{A}_{\vec{k}}$ and $\mat{B}_{\vec{k}}$ are given by
\begin{subalign}
  A^{\alpha\alpha'}_{\vec{k}} &= S \left(\mathcal{J}^{+-}_{\vec{k},\alpha\alpha'}
  -\delta_{\alpha\alpha'} \sum_{\mu} \mathcal{J}^{00}_{\vec{0},\alpha\mu}\right), \\
  B^{\alpha\alpha'}_{\vec{k}} &= S\mathcal{J}^{++}_{\vec{k},\alpha\alpha'},
\end{subalign}
where we have defined the Fourier transforms of the local
exchange matrices as
\begin{equation}
  \label{eq:lswt:local}
\mathcal{J}^{\mu\mu'}_{\vec{k},\alpha\alpha'} \equiv \sum_{\vec{\delta}} \mathcal{J}^{\mu\mu'}_{\vec{\delta},\alpha\alpha'}e^{i\vec{k}\cdot\vec{\delta}} .
\end{equation}
The linear spin-wave Hamiltonian [Eq.~(\ref{eq:lswt:ham})] can be diagonalized by a Bogoliubov transformation. To do
this one diagonalizes the modified matrix~\cite{blaizot:1986}
\begin{align}
  \label{eq:lswt:bogo}
    \left(
    \begin{array}{cc}
      \mat{A}_{\vec{k}} & \mat{B}_{\vec{k}} \\
      -\cb{\mat{B}}_{-\vec{k}} & -\cb{\mat{A}}_{-\vec{k}}
    \end{array}
  \right) \equiv   \mat{\sigma}_3 \mat{M}_{\vec{k}} 
\end{align}
This yields pairs of eigenvectors $\vec{V}_{\vec{k}\alpha}$ and
$\vec{W}_{-\vec{k},\alpha} = \mat{\sigma}_1
\cb{\vec{V}}_{-\vec{k}\alpha}$ with eigenvalues
$+\epsilon_{\vec{k}\alpha}$ and $-\epsilon_{-\vec{k},\alpha}$. These
vectors can be normalized such that~\cite{blaizot:1986}
\begin{align}
  \label{eq:lswt:norm}
  \h{\vec{V}}_{\vec{k}\alpha} \mat{\sigma}_3 \nh{\vec{V}}_{\vec{k}\alpha'}
  &= +\delta_{\alpha\alpha'},
  &
    \h{\vec{W}}_{-\vec{k}\alpha} \mat{\sigma}_3 \nh{\vec{W}}_{-\vec{k}\alpha'}
  &= -\delta_{\alpha\alpha'},
  &
    \h{\vec{W}}_{-\vec{k}\alpha} \mat{\sigma}_3 \nh{\vec{V}}_{\vec{k}\alpha'}
  &= 0.
\end{align}
One can then write the Hamiltonian in terms of diagonalized bosons, $\gamma_{\vec{k}\alpha}$, as~\cite{blaizot:1986}
\begin{align}
  \label{eq:lswt:diag}
  H
  \equiv
    N S(S+1)\epsilon_{\rm cl} + N  \epsilon_{\rm qu} +
  \sum_{\vec{k}\alpha}\epsilon_{\vec{k}\alpha} \h{\gamma}_{\vec{k}\alpha} \nh{\gamma}_{\vec{k}\alpha}  + O(S^{1/2}),
\end{align}
where we have identified the energy per site from quantum zero-point
motion, $\epsilon_{\rm qu}$, as
\begin{align}
  \label{eq:lswt:quantum}
  \epsilon_{\rm qu} &\equiv
                      \frac{1}{2N}\sum_{\vec{k}\alpha} \epsilon_{\vec{k}\alpha}
\end{align}
\subsubsection{Spin Waves at $O(1/S)$}
\begin{figure}
    \includegraphics[width=\textwidth]{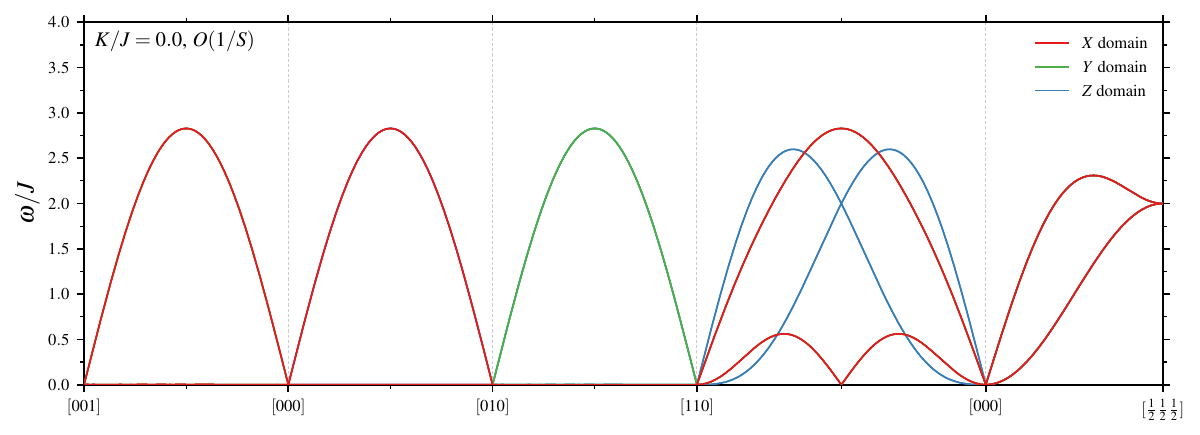}
    \includegraphics[width=\textwidth]{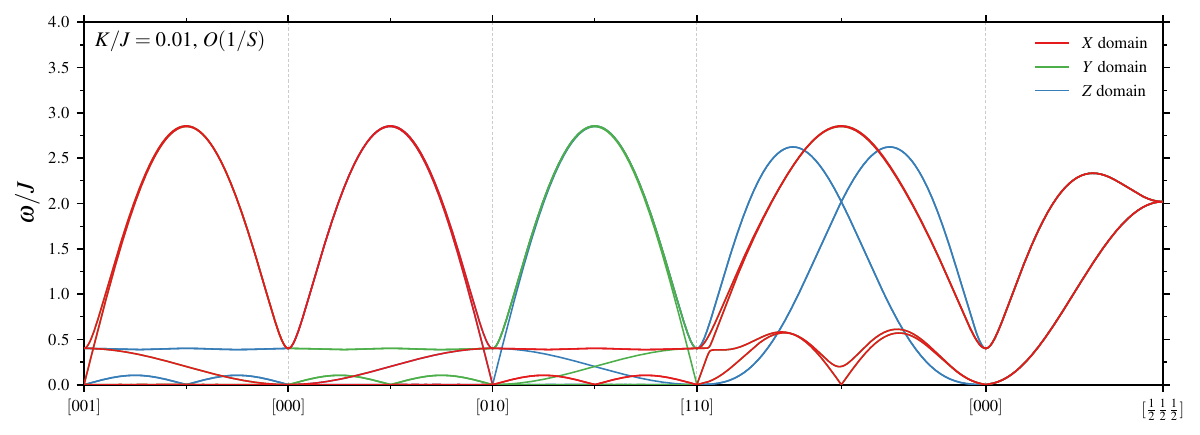}    
    \includegraphics[width=\textwidth]{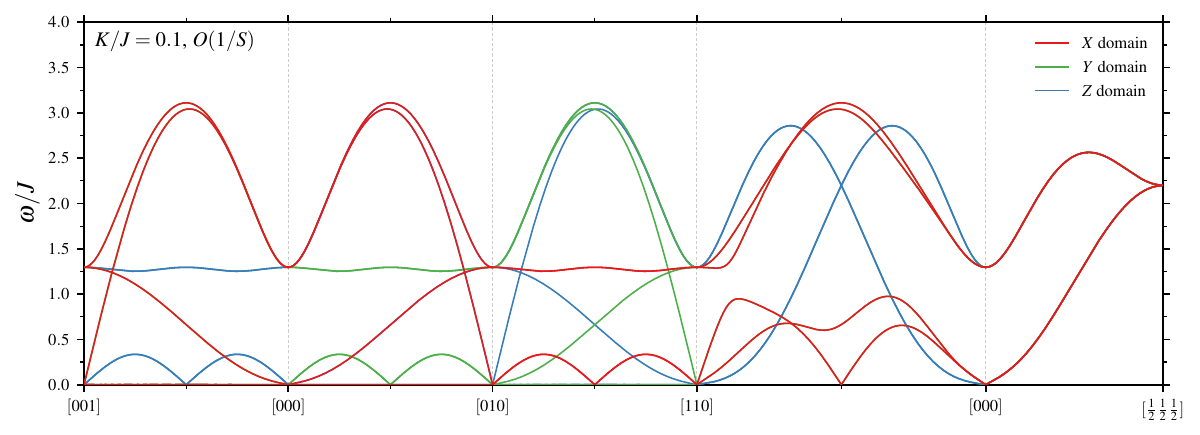}    
    \caption{
    Linear spin-wave spectrum of the $S=1/2$ nearest-neighbor model [Eq.~(\ref{supp:eq:jkg})] in the type III antiferromagnet along a high-symmetry path 
    with $J>0$ and $\Gamma=0$.
    Spectra for several values of $K/J$ and for three symmetry related domains are shown. 
    We see that some of the line nodes present when $K/J=0$ are lifted by finite $K/J$, but those
    that are perpendicular to the stacking direction remain, along with several pseudo-Goldstone modes.
    \label{supp:fig:lswt-hk}
    }
\end{figure}
The type III ordering observed in \KIC{} can be described using a four-site magnetic unit cell using the same four atoms of the conventional cubic cell. We write
\begin{align*}
    \vec{r}_1 &= \vec{0} &
    \vec{r}_2 &= \vec{a}_1 &
    \vec{r}_3 &= \vec{a}_2 &
    \vec{r}_4 &= \vec{a}_3 
\end{align*}
where $\vec{a}_1 = a(\vhat{y}+\vhat{z})/2$, $\vec{a}_2 = a(\vhat{x}+\vhat{z})/2$ and $\vec{a}_3=(\vhat{x}+\vhat{y})/2$ are the usual primitive lattice vectors. 
The type III domains can be described by ordering directions on each sublattice $\vhat{z}_1$, $\vhat{z}_2$, $\vhat{z}_3$, $\vhat{z}_4$ and new lattice translations $\vec{A}_1$, $\vec{A}_2$, $\vec{A}_3$, as given below
\begin{center}
$$
\begin{array}{c|cccc|ccc}
    \textrm{Domain} & \vhat{z}_1 &\vhat{z}_2 &\vhat{z}_3 &\vhat{z}_4 & \vec{A}_1 & \vec{A}_2 & \vec{A}_3 \\
    \hline
    X & \pm\vhat{x} & \mp\vhat{x} & \pm\vhat{x} & \mp\vhat{x} & a\vhat{x}+\vec{a}_1 & a\vhat{y} & a\vhat{z} \\
    & \pm\vhat{x} & \mp\vhat{x} & \mp\vhat{x} & \pm\vhat{x} & a\vhat{x}+\vec{a}_1 & a\vhat{y} & a\vhat{z} \\
    Y & \pm\vhat{y} & \mp\vhat{y} & \mp\vhat{y} & \pm\vhat{y} & a\vhat{x} & a\vhat{y}+\vec{a}_2 & a\vhat{z} \\
    & \pm\vhat{y} & \pm\vhat{y} & \mp\vhat{y} & \mp\vhat{y} & a\vhat{x} & a\vhat{y}+\vec{a}_2 & a\vhat{z} \\
    Z & \pm\vhat{z} & \pm\vhat{z} & \mp\vhat{z} & \mp\vhat{z} & a\vhat{x} & a\vhat{y} & a\vhat{z}+\vec{a}_3 \\ 
    & \pm\vhat{z} & \mp\vhat{z} & \pm\vhat{z} & \mp\vhat{z} & a\vhat{x} & a\vhat{y} & a\vhat{z}+\vec{a}_3 \\         
\end{array}
$$
\end{center}
Note that we have provided four sign structures for the $\vhat{z}_{\alpha}$ of each of the $X$, $Y$ and $Z$ domains, giving in total twelve domains (as expected). The four sign structures correspond to different type III states related by translational or time-reversal symmetries (leaving the spectrum and scattering intensity unchanged), while the $X$, $Y$ and $Z$ families are related by a three-fold rotation.

For each choice of domain and sign structure we can construct the Fourier transforms of the local exchange matrices,
$\mathcal{J}^{\mu\nu}_{\vec{k},\alpha\alpha'}$, as outlined in Sec.~\ref{supp:sec:lswt} and from those the $\mat{A}_{\vec{k}}$ and $\mat{B}_{\vec{k}}$ matrices. Each of these are four by four matrices, leading to a final eight by eight eigenproblem encoded in $\mat{M}_{\vec{k}}$ to determine the spin-wave energies and intensities. Practically, we tabulate a list of nearest-neighbor bonds originating within our unit cell and construct these matrices numerically for each wave-vector of interest.

Results for the Heisenberg-Kitaev limit ($\Gamma/J=0$) at $S=1/2$ are shown in Fig.~\ref{supp:fig:lswt-hk}, with three inequivalent domains indicated. We see that for $K/J=0$ one has nodal lines along the $[1,0,0]$, $[0,1,0]$ and $[0,0,1]$ directions (or equivalents). As $K/J$ is rendered finite we see that the nodal lines perpendicular to the stacking are preserved, while those
along the stacking direction are lifted. Further, in addition to these
nodal lines we see several pseudo-Goldstone modes at wave-vectors characteristic of the type III order (such as $(1, 0.5, 0)$ and equivalents). The presence of finite $\Gamma$ with $|\Gamma|<K/2$ does affect the linear spin-wave spectrum, but only away from high-symmetry lines (for details see Sec.~\ref{supp:sec:effect-of-gamma}), and the nodal lines and pseudo-Goldstone modes are preserved. We thus see the linear spin-wave spectrum is qualitatively different than what is observed experimentally due to the presence of a large number of zero modes.

\begin{figure}
    \includegraphics[width=\textwidth]{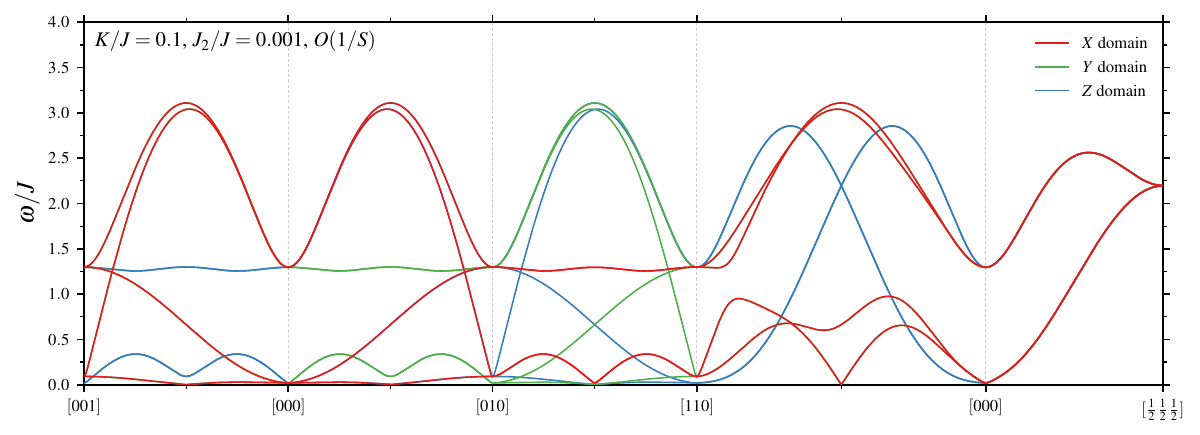}
    \includegraphics[width=\textwidth]{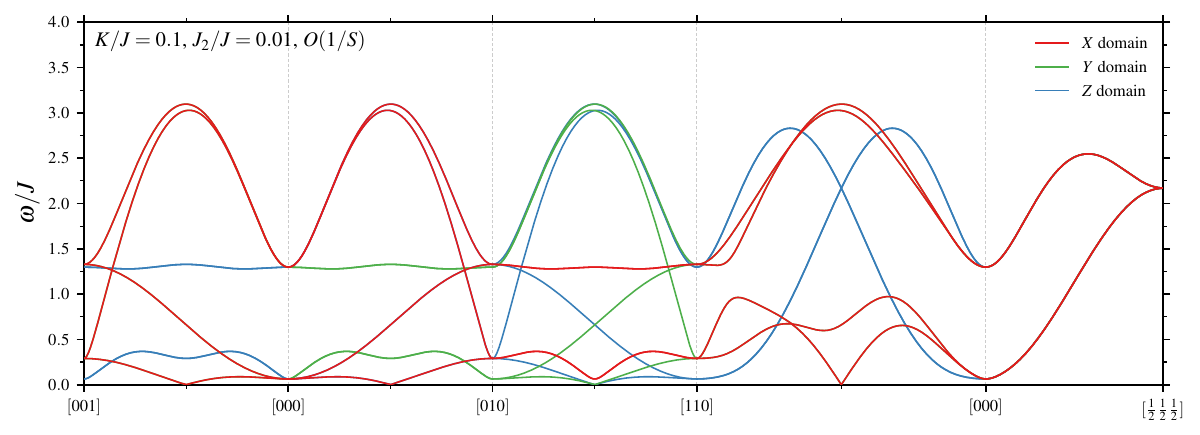}
    \includegraphics[width=\textwidth]{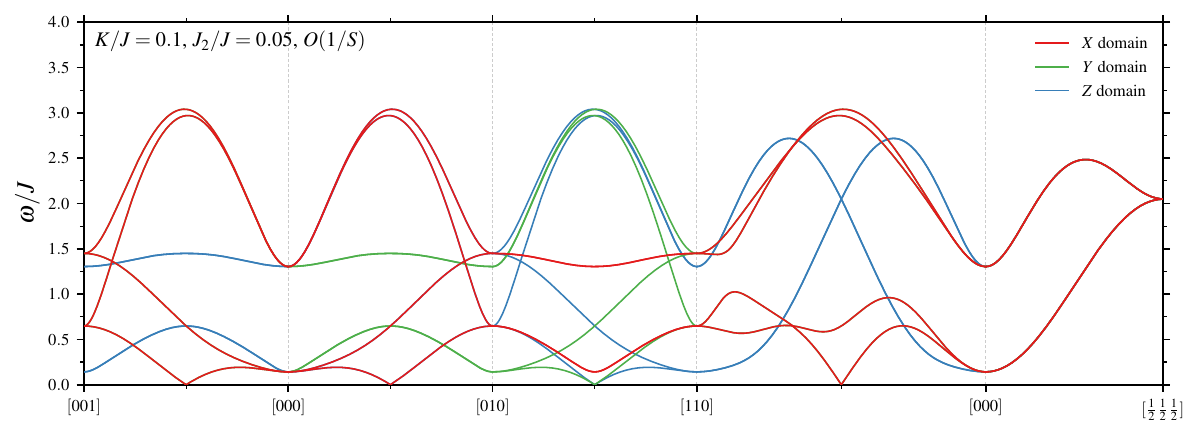}
    \caption{
    Linear spin-wave spectrum of the $S=1/2$ nearest-neighbor model in the type III antiferromagnet along a high-symmetry path 
    with $J>0$, $\Gamma=0$ and finite $J_2/J>0$.
    Spectra for several values of $J_2/J$ with $K/J=0.1$ and for three symmetry related domains are shown. 
    We see that some of the line nodes present when $J_2/J=0$ are lifted by finite $J_2/J$, but several pseudo-Goldstone modes remain.
    \label{supp:fig:lswt-hk-j2}
    }
\end{figure}

As the accidental degeneracy that gives rise to these zero modes can be lifted by a finite second neighbor exchange $J_2$, we should expect that the nodal lines will be gapped for $J_2/J>0$ (stabilizing the type III order). The effect of finite $J_2/J$ is shown in Fig.~\ref{supp:fig:lswt-hk-j2} for several values of $J_2/J$. We see that indeed the remaining nodal lines are lifted, but several of the pseudo-Goldstone modes associated with the type III ordering wave-vectors remain. This is a consequence of the accidental degeneracy of the of the three type III domains that is not lifted by finite $J_2/J$. As discussed in Sec.~\ref{supp:sec:effect-of-finite-j2}, this degeneracy is robust and is not lifted by \emph{any} bilinear exchanges consistent with the symmetry of the model. We thus expect these pseudo-Goldstone modes to be persistent at the linear spin-wave level even if additional anisotropic exchanges (e.g. a second neighbor Kitaev coupling) or further neighbor interactions are included. As with the Heisenberg-Kitaev model itself, the presence of these low-lying bands and gapless pseudo-Goldstone modes disagrees qualitatively with what is observed experimentally.

\subsubsection{Order by Quantum Disorder}
In the Heisenberg limit, quantum fluctuations at $O(1/S)$ are expected to select the collinear type III ordered from the stacked states described in Sec.~\ref{supp:sec:ground-state-manifold}. This was explored in Ref.~[\onlinecite{schick:2020}] for type I and III states, as well as a family of incommensurate spirals. We have further confirmed that at $O(1/S)$ that among all collinear stacked states up to period-12 in the stacking direction that type III remains the minimum. We have also confirmed that non-collinear states obtained by continuously mixing different type I states, different type III states as well as type I and type III states do not yield a different selection. These results hold true at finite $K/J$ and for small $\Gamma$ as well.

Superficially this would suggest order-by-quantum-disorder selection at $O(1/S)$ could be responsible for the appearance of type III order in \KIC{}. However, as shown in Ref.~\onlinecite{schick:2022}, $O(1/S^2)$ and higher corrections likely change this result in the Heisenberg limit. How this result extends to finite $K/J$ or finite $\Gamma/J$ is unclear. Calculations at $O(1/S^2)$ following the strategy of Ref.~\cite{schick:2022} are inconclusive when $K/J \neq 0$, with the type I state developing negative energy modes at $S=1/2$. We leave exploration of the $O(1/S^2)$ order-by-quantum-disorder selection in the degenerate phase of the nearest-neighbor model [Eq.~(\ref{supp:eq:jkg})] to future work.

\subsubsection{Dynamical Structure Factor}
\label{supp:sec:int-lswt}
The inelastic neutron scattering intensity per spin is determined by the spin-spin correlation
function
$$
\mathcal{S}_{\mu\nu}(\vec{k},\omega) \equiv
\frac{1}{2\pi N} \int dt e^{-i\omega t}
\avg{M^\mu_{-\vec{k}} M^\nu_{\vec{k}}(t)}
$$
where $\vec{M}_{\vec{k}} \equiv g\mu_B \sum_i e^{-i\vec{k}\cdot \vec{r}_i} \vec{S}_i $ is the magnetization
operator at wave-vector $\vec{k}$ and $N$ is the total number of spins. Explicitly the intensity per spin is given by
\begin{align*}
    I(\vec{k},\omega) &=
    \frac{k_f}{k_i} \left(\frac{\gamma r_0}{2\mu_B}\right)^2 F(k)^2 \sum_{\mu\nu} \left(\delta_{\mu\nu}-\hat{k}_{\mu}\hat{k}_{\nu}\right) \mathcal{S}_{\mu\nu}(\vec{k},\omega)
\end{align*}
where $r_0 = \mu_0 e^2/(4\pi m_e) \approx 2.818 \cdot 10^{-15}$ m is the classical electron radius, $\mu_B$ is the Bohr magneton,
$\gamma = 1.913$, $F(k)$ is the magnetic form factor and $k_i$, $k_f$ are the incoming and outgoing neutron wave-vectors. It is useful to write $(\gamma r_0)^2 \approx 0.291$ barns where $1$ barn $= 10^{-28}$ m.

Within linear spin-wave theory the dynamical structure factor can be expressed in terms of the transverse-transverse part of the spin-spin correlation function. At zero temperature the dynamical structure factor with linear spin-wave theory takes the form~\cite{fishman:2018}
$$
\mathcal{S}_{\mu\nu}(\vec{k},\omega)=\sum_{\vec{k},n}
W^{\mu\nu}_{\vec{k},n}
\delta(\omega - \omega_{\vec{k},n})
$$
where we defined the weights for each spin-wave mode as
$
W^{\mu\nu}_{\vec{k},n}
\equiv \left(S \mu_B g^2/N_s\right) \Phi^{n,\mu}_{\vec{k}} \bar{\Phi}^{n,\mu'}_{\vec{k}}
$
where $N_s$ is the number magnetic sublattices.
The quantities $\Phi^{n,\mu}_{\vec{k}}$ are defined as
$$
\Phi^{n,\mu}_{\vec{k}} \equiv \sum_{\alpha} e^{-i\vec{k}\cdot \vec{r}_{\alpha}}\left(
X^{\alpha}_{\vec{k},n} \hat{e}^{\mu}_{\alpha,-}+Y^{\alpha}_{\vec{k},n} \hat{e}^{\mu}_{\alpha,+}
\right)
$$
The vectors $\mat{X}_{\vec{k}}$ and $\mat{Y}_{\vec{k}}$ are blocks of the eigenvectors
of $\mat{M}_{\vec{k}}$ with $\vec{V}_{\vec{k},n} = (\vec{X}_{\vec{k},n},\vec{Y}_{\vec{k},n})$.

Note that the gap observed experimentally is much larger than the sample temperature, $\Delta \sim 0.65$ meV compared to $T \sim 0.25$ K and so we are justified in taking the zero temperature limit. Corrections due finite temperature effects will be bounded by the Bose factor $n_B(\Delta)$, which is entirely negligible at $0.25$ K.

\subsection{Non-Linear Spin Wave Theory}
\label{supp:sec:nlswt}
Going to next order in $1/S$, one obtains a more complex Hamiltonian for the magnon excitations. This can be
written as a sum of the usual two-magnon terms, as well as generally three- and four-magnon interactions. Following the insights from Sec.~\ref{supp:sec:lswt} we will consider only the case where $\Gamma=0$. With this restriction the three-magnon interaction terms vanish and the spin-wave interactions simplify considerably. One obtains
\begin{equation}
  H = N S(S+1) \epsilon_{\rm cl} + H_2  + H_4
\end{equation}
where we define the individual pieces in symmetrized form as
\begin{subalign}
  {H}_2 &= \frac{1}{2}\sum_{\alpha\beta}\sum_{\vec{k}} \left[
    {A}_{\vec{k}}^{\alpha\beta} \h{a}_{\vec{k}\alpha}\nh{a}_{\vec{k}\beta} +
    {A}_{-\vec{k}}^{\beta\alpha} \nh{a}_{-\vec{k}\alpha}\h{a}_{-\vec{k}\beta} +
           \left(
          {B}^{\alpha\beta}_{\vec{k}}\h{a}_{\vec{k}\alpha}\h{a}_{-\vec{k}\beta} +
          \cb{B}^{\alpha\beta}_{\vec{k}}\nh{a}_{-\vec{k}\beta}\nh{a}_{\vec{k}\alpha}\right)
          \right],\nonumber \\
  {H}_4 &= \frac{1}{N_c}\sum_{\alpha\beta\mu\nu} \sum_{\vec{k}\vec{k}'\vec{q}}\left[
                \frac{1}{(2!)^2} {V}_{\vec{k}\vec{k}[\vec{q}]}^{\alpha\beta\mu\nu} \h{a}_{\vec{k}+\vec{q},\alpha}\h{a}_{\vec{k}'-\vec{q},\beta} \nh{a}_{\vec{k}'\mu} \nh{a}_{\vec{k}\nu} +
                \frac{1}{3!} \left(
                {D}_{\vec{k}\vec{k}'\vec{q}}^{\alpha\beta\mu\nu}\h{a}_{\vec{k}\alpha}\h{a}_{\vec{k}'\beta} \h{a}_{\vec{q}\mu} \nh{a}_{\vec{k}+\vec{k}'+\vec{q},\nu} +\hc
                \right)
                \right].\nonumber
\end{subalign}
In terms of the local exchange matrices [Eq.~(\ref{eq:lswt:exchange})] one can write            
\begin{subalign}
  {A}_{\vec{k}}^{\alpha\beta}
  &=
    S\left(\mathcal{J}^{+-}_{\vec{k},\alpha\beta} - \delta_{\alpha\beta}\sum_\mu \mathcal{J}^{00}_{\vec{0},\alpha\mu}\right), \\
  {B}_{\vec{k}}^{\alpha\beta}
  &=
    S\mathcal{J}^{++}_{\vec{k},\alpha\beta},  \\
  {V}^{\alpha\beta\mu\nu}_{\vec{k}\vec{k}'[\vec{q}]}
  &=
    \left(\delta_{\alpha\mu}\delta_{\beta\nu}\mathcal{J}^{00}_{\vec{k}-\vec{k}'+\vec{q},\alpha\beta}+
    \delta_{\alpha\nu}\delta_{\beta\mu}\mathcal{J}^{00}_{\vec{q},\alpha\beta}\right)-
    \left(
    \delta_{\mu\nu}\delta_{\mu\beta}\mathcal{J}^{+-}_{\vec{k}+\vec{q},\alpha\nu}+
    \delta_{\alpha\beta}\delta_{\alpha\mu}\mathcal{J}^{+-}_{\vec{k},\alpha\nu}
    \right), \\
  {D}^{\alpha\beta\mu\nu}_{\vec{k}\vec{k}'\vec{q}}
  &=
   -\frac{3}{4} \left(
    \delta_{\alpha\mu}\delta_{\alpha\nu} \mathcal{J}^{++}_{\vec{k}',\beta\alpha}+
    \delta_{\mu\beta}\delta_{\nu\beta} \mathcal{J}^{++}_{\vec{k},\alpha\beta}\right),
\end{subalign}
where the four-magnon vertices have been left unsymmetrized for brevity. At leading order in perturbation theory, these magnon interaction terms renormalize the linear spectrum, giving corrections to $\mat{A}_{\vec{k}}$ and $\mat{B}_{\vec{k}}$
\begin{subalign}
\Delta A^{\alpha\beta}_{\vec{k}} &= \frac{1}{N_c} \sum_{\vec{q}} \sum_{\mu\nu}\left[
  V^{\alpha \mu \nu \beta}_{\vec{k}\vec{q}[\vec{0}]}\avg{\h{a}_{\vec{q}\mu}\nh{a}_{\vec{q}\nu}}_0 +
  \frac{1}{2}\left(
  D^{\alpha \mu \nu \beta}_{\vec{k},-\vec{q},\vec{q}}\avg{\h{a}_{-\vec{q}\mu}\h{a}_{\vec{q}\nu}}_0+
  \cb{D}^{\beta\mu\nu\alpha}_{\vec{k},\vec{q},-\vec{q}}\avg{\nh{a}_{\vec{q}\mu}\nh{a}_{-\vec{q}\nu}}_0
  \right)
  \right], \nonumber \\
  \Delta B^{\alpha\beta}_{\vec{k}} &=
\frac{1}{N_c} \sum_{\vec{q}} \sum_{\mu\nu}\left[
  D^{\mu\alpha\beta\nu}_{\vec{q},\vec{k},-\vec{k}} \avg{\h{a}_{\vec{q}\mu}\nh{a}_{\vec{q}\nu}}_0 +
  \frac{1}{2}
    V^{\alpha\beta \nu \mu }_{\vec{q},-\vec{q},[\vec{k}-\vec{q}]}\avg{\nh{a}_{\vec{q}\mu}\nh{a}_{-\vec{q}\nu}}_0 
  \right].  \nonumber 
\end{subalign}
where $\avg{\dots}_0$ is an average with respect to the linear spin-wave Hamiltonian, $H_2$.
\subsubsection{Self-consistent Spin Waves}
\label{supp:sec:self-consistent}
Due to the large number of zero modes present in the linear spin-wave spectrum, divergences in the interactions can make the leading perturbative result unreliable~\cite{schick:2020}. One way to resolve these divergences is by including the interaction terms self-consistently~\cite{schick:2022}. With
this in mind we consider a adding and subtracting a quadratic piece to
the Hamiltonian, writing
\begin{equation}
  H_2 + H_4 = \left(H_2+\delta H_2\right) + \left(H_4 - \delta H_2\right).
\end{equation}
where we define $\delta H_2$ as 
\begin{equation}
  \delta H_2 \equiv
  \sum_{\alpha\beta}\sum_{\vec{k}} \left[
    \delta {A}_{\vec{k}}^{\alpha\beta} \h{a}_{\vec{k}\alpha}\nh{a}_{\vec{k}\beta} +
    \frac{1}{2} \left(
      \delta {B}^{\alpha\beta}_{\vec{k}}\h{a}_{\vec{k}\alpha}\h{a}_{-\vec{k}\beta} +
      \delta \cb{B}^{\alpha\beta}_{\vec{k}}\nh{a}_{-\vec{k}\beta}\nh{a}_{\vec{k}\alpha}\right)
  \right]  .
\end{equation}
 We now consider $H_2
+ \delta H_2$ as the linear spin-wave problem, evaluating the energies
and eigenvectors with respect to this Hamiltonian. We treat $\delta
H_2$ perturbatively; it appears once at leading order and thus simply produces an additive contribution to $\mat{A}_{\vec{k}}$ and $\mat{B}_{\vec{k}}$.
We define our self-consistent mean-field theory by choosing the
perturbations $\delta \mat{A}_{\vec{k}}$ and $\delta
\mat{B}_{\vec{k}}$ so that they \emph{cancel} the corrections produced by the interactions. This can be done via an iterative process
\newcommand{\scavg}[1]{\textrm{MF}}
\begin{subalign}
\delta A^{\alpha\beta}_{\vec{k}} &\leftarrow \frac{1}{N_c} \sum_{\vec{q}} \sum_{\mu\nu}\left[
  V^{\alpha \mu \nu \beta}_{\vec{k}\vec{q}[\vec{0}]}\avg{\h{a}_{\vec{q}\mu}\nh{a}_{\vec{q}\nu}}_{\scavg{}} +
  \frac{1}{2}\left(
  D^{\alpha \mu \nu \beta}_{\vec{k},-\vec{q},\vec{q}}\avg{\h{a}_{-\vec{q}\mu}\h{a}_{\vec{q}\nu}}_{\scavg{}}+
  \cb{D}^{\beta\mu\nu\alpha}_{\vec{k},\vec{q},-\vec{q}}\avg{\nh{a}_{\vec{q}\mu}\nh{a}_{-\vec{q}\nu}}_{\scavg{}}
  \right)
  \right],\nonumber \\
  \delta B^{\alpha\beta}_{\vec{k}} &\leftarrow
\frac{1}{N_c} \sum_{\vec{q}} \sum_{\mu\nu}\left[
  D^{\mu\alpha\beta\nu}_{\vec{q},\vec{k},-\vec{k}} \avg{\h{a}_{\vec{q}\mu}\nh{a}_{\vec{q}\nu}}_{\scavg{}} +
  \frac{1}{2}
    V^{\alpha\beta \nu \mu }_{\vec{q},-\vec{q},[\vec{k}-\vec{q}]}\avg{\nh{a}_{\vec{q}\mu}\nh{a}_{-\vec{q}\nu}}_{\scavg{}}
  \right]. \nonumber
\end{subalign}
where $\avg{\dots}_{\scavg{}}$ is evaluated with respect to the shifted Hamiltonian $H_2 +\delta H_2$ and depends on the current values of 
$\delta \mat{A}_{\vec{k}}$ and $\delta\mat{B}_{\vec{k}}$. Once converged, one has an effective quadratic model of the spin-wave spectrum, with interaction effects encoded
in $\delta \mat{A}_{\vec{k}}$ and $\delta\mat{B}_{\vec{k}}$.

\subsubsection{Spin Waves at $O(1/S^2)$}
We follow the self-consistent scheme discussed in Sec.~\ref{supp:sec:self-consistent}. As for the linear spin-wave case, we tabulate a list of nearest-neighbor bonds without our unit cell and construct the matrices and interaction vertices numerically for each wave-vector needed in the self-consistent sums.

Practically, we iterate this loop for each wave-vector until the corrections have stopped changing to an absolute tolerance of size $10^{-8}$ in all matrix elements.
To resolve some of the divergences due to zero modes present when interactions are absent, the $\delta \mat{A}_{\vec{k}}$ and $\delta\mat{B}_{\vec{k}}$ matrices were initialized to non-zero values at the start of the iterative loop with $\delta \mat{A}_{\vec{k}}=\mu \mat{1}$ and $\delta\mat{B}_{\vec{k}}=\mat{0}$ with $\mu>0$. We have confirmed that the converged result is independent of the precise choice of $\mu$. 

The resulting $\delta \mat{A}_{\vec{k}}$ and $\delta\mat{B}_{\vec{k}}$ yield a quadratic spin-wave Hamiltonian with $\mat{M}^{\rm eff}_{\vec{k}} = \mat{M}_{\vec{k}} + \delta \mat{M}_{\vec{k}}$ which can be diagonalized as in Sec.~\ref{supp:sec:lswt} to yield spin-wave energies and wave-functions.

\begin{figure}
    \includegraphics[width=\textwidth]{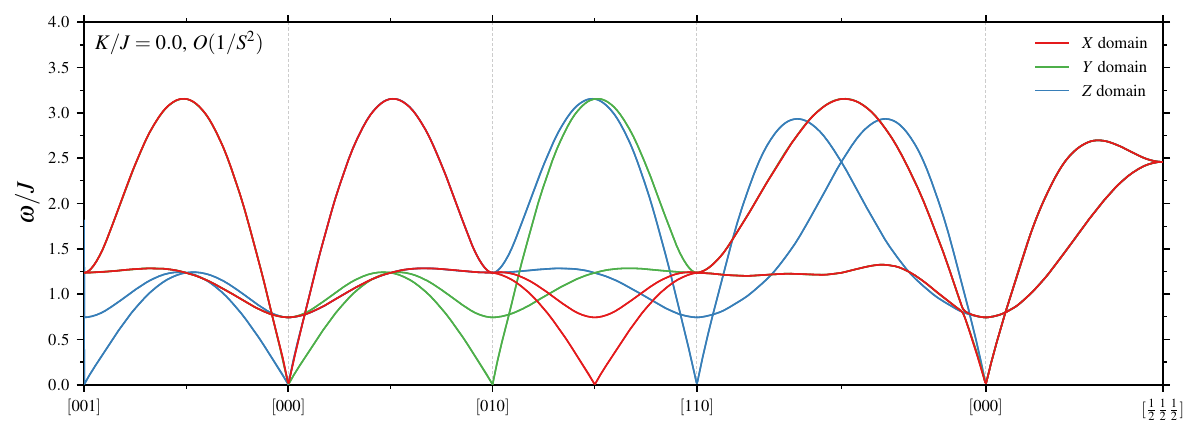}
    \includegraphics[width=\textwidth]{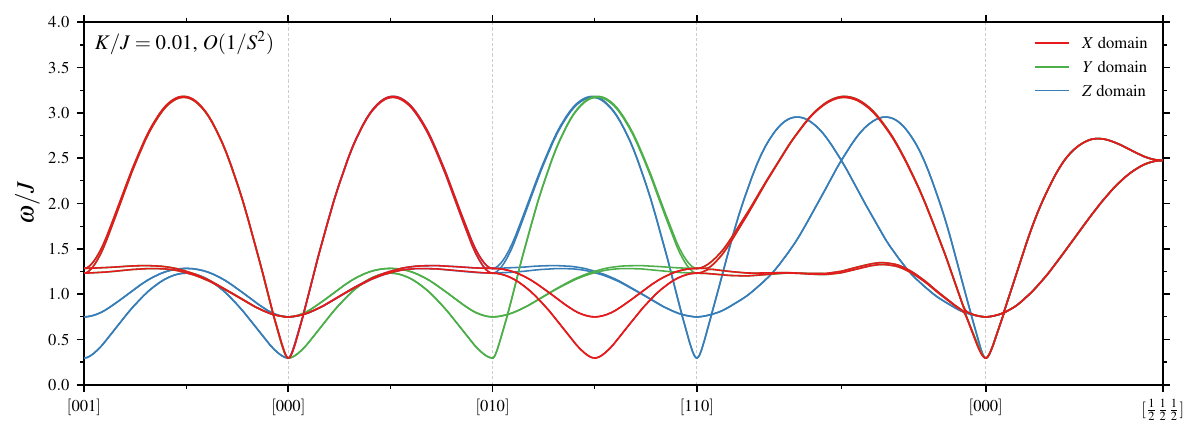}    
    \includegraphics[width=\textwidth]{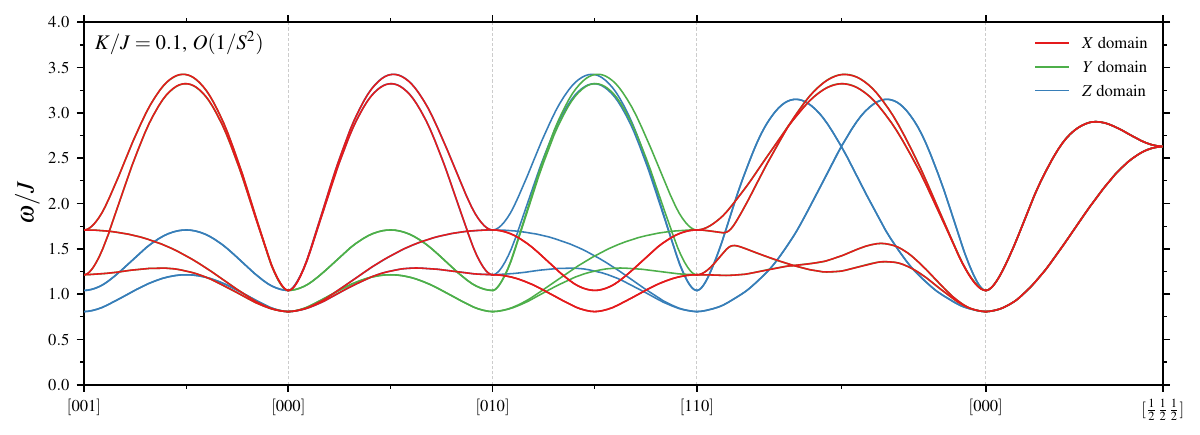}
    \caption{
    Self-consistent non-linear spin-wave spectrum of the $S=1/2$ nearest-neighbor model [Eq.~(\ref{supp:eq:jkg})] in the type III antiferromagnet along a high-symmetry path 
    with $J>0$ and $\Gamma=0$.
    Spectra for several values of $K/J$ and for three symmetry related domains are shown. 
    We see that the remaining line nodes present when $K/J=0$ in linear spin-wave theory are lifted by interaction corrections,
    as are all remaining pseudo-Goldstone modes.    
    \label{supp:fig:sc-nlswt-hk}
    }
\end{figure}
\begin{figure}
    \includegraphics[width=\textwidth]{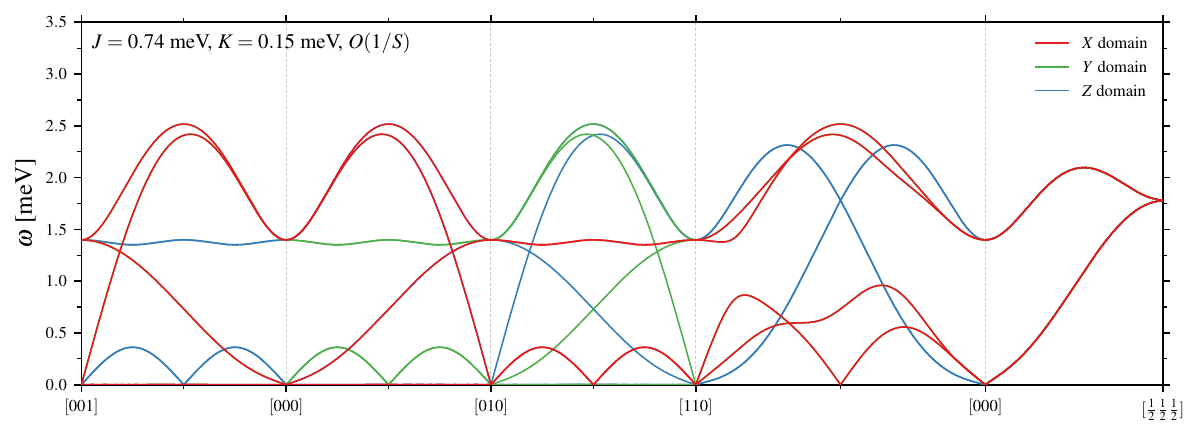}
    \includegraphics[width=\textwidth]{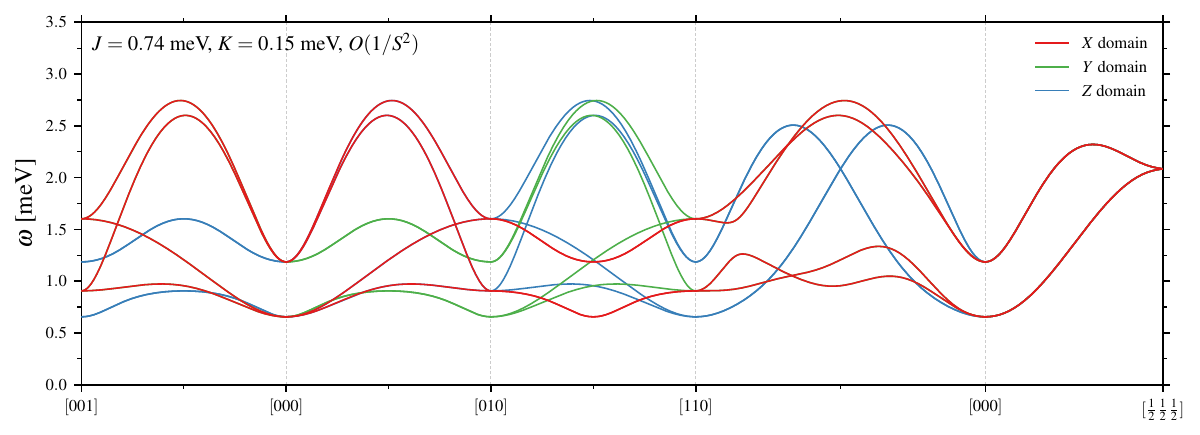} 
    \caption{
    Linear and self-consistent non-linear spin-wave spectrum of the $S=1/2$ nearest-neighbor model [Eq.~(\ref{supp:eq:jkg})] in the type III antiferromagnet along a high-symmetry path 
    for the best fit parameters $J=0.74$ meV and $K=0.15$ meV. Three symmetry related domains are shown. We see that the presence of spin-wave interactions introduces a large gap to all of the nodal lines and pseudo-Goldstone modes present in the linear spin-wave spectrum.    
    \label{supp:fig:swt-best-fit}
    }
\end{figure}

\subsubsection{Corrections to the Dynamical Structure Factor}
\label{supp:sec:corrections-to-dyn-sf}
The higher order terms in the Holstein-Primakoff expansion also induce corrections to the spin-spin correlation functions that appear in the dynamical structure factor. When magnon decay is forbidden, this generally involves modifications to the intensities from the transverse spin-spin correlators, as well as contributions to the longitudinal spin-spin correlators from the two-magnon continuum~\cite{mourigal:2013}. 

The renormalization of the transverse part 
depends on the normal and anomalous on-site averages~\cite{zhitomirsky:2013,mourigal:2013}
\begin{align*}
    n_{\alpha} &\equiv \avg{\h{a}_{\vec{r}\alpha} a_{\vec{r}\alpha}}_0 &
    \delta_{\alpha} &\equiv \avg{{a}_{\vec{r}\alpha} a_{\vec{r}\alpha}}_0
\end{align*}
For the nearest-neighbor model with $\Gamma=0$ we have that $n_{\alpha} \equiv n$ which is independent of sublattice and $\delta_{\alpha}=0$. In this scenario the dynamical structure factor in the self-consistent theory can then be computed from $H_2+\delta H_2$ as in standard linear-spin-wave theory (see Sec.~\ref{supp:sec:int-lswt}), save with the an overall intensity correction 
$$
I(\vec{k},\omega) \rightarrow \left(1-\frac{n}{2S}\right) I(\vec{k},\omega)
$$
We have included this correction in our comparisons to our experimental data. We note that for our best fit parameters the size of $n$ is not particular large, $n\sim 0.1$, and this correction only amounts to a $10\%$ reduction in scattering intensity.

The contribution from the two-magnon continuum was found to be insignificant in our self-consistent spin-wave theory, with the spectral weight amounting to a small fraction of the contribution of the transverse parts. We have therefore not included these contributions in our comparisons to the experimental data. Further, we note that these contributions would only modify the intensities for $\omega \geq 2\Delta \sim 1.3 \meV$ where $\Delta \sim 0.65 \meV$ is the one-magnon gap. This intensity also only increases slowly from that minimum, following roughly the two-magnon density of states.

\subsection{Effect of Tetragonal Distortion}
If the symmetry of the crystal structure is lowered from cubic to tetragonal, additional exchange interactions 
are allowed in the spin Hamiltonian. For illustration we consider the allowed exchange interactions in $I4/mmm$ (\#139) which is one of the larger tetragonal subgroups of $Fm\bar{3}m$ (\#225). 
A standard analysis of the space group symmetry (taking $\vhat{z}$ to be the tetragonal axis) yields 
\begin{align}
\label{supp:eq:tetragonal-model}
H = \sum_{\avg{ij}_{z}}&\left[
    J_{\perp}\vec{S}_i\cdot \vec{S}_j+
                         \left(K_{\perp}+A_{\perp} \right)S^{z}_i S^{z}_j +
    (-1)^{\sigma^{xy}_{ij}} \Gamma_{\perp} \left(S^{x}_i S^{y}_j + S^{y}_i S^{x}_j\right)
    \right] + \nonumber \\
\sum_{\gamma=x,y}\sum_{\avg{ij}_{\gamma}}&\left[
    J \vec{S}_i\cdot \vec{S}_j+
    K S^{\gamma}_i S^{\gamma}_j +
    (-1)^{\sigma^{\alpha\beta}_{ij}} \Gamma \left(S^{\alpha}_i S^{\beta}_j + S^{\alpha}_i S^{\beta}_j\right)
                                           + A_{\perp} S^z_i S^z_j
                                           \right]
\end{align}
This model includes different $J$, $K$ and $\Gamma$ interactions in the perpendicular to the tetragonal axis, as well as global XXZ anisotropy encoded in $A_{\perp}$. Given a small lattice distortion we should loosely
expect each of these terms to be proportional to some component of the strain $\epsilon$, as well
as a spin-lattice coupling constant. The presence of these anisotropies breaks the continuous degeneracy between the type III states and other stacked classical ground states.
\subsubsection{Effect on Magnon Gap in Type-III State}
At the linear spin-wave level the magnon energies at the high-symmetry wave-vectors
$(1,0,0)$, $(0,1,0)$ and $(0,0,1)$ for the $Z$-domain of the type-III order gives two distinct energies for each wave-vector 
\begin{align*}
  \omega_{[001],1} &= 2\sqrt{(K_{\perp}+A_{\perp})(2(J_{\perp}-J)+K_{\perp}-K+A_{\perp})},\\
  \omega_{[001],2} &= 2\sqrt{(K_{\perp}+A_{\perp})(2(J_{\perp}+J)+K_{\perp}+K+A_{\perp})},\\
  \omega_{[100],1} = \omega_{[010],1} &= 2\sqrt{(K_{\perp}-K+A_{\perp})(2J_{\perp}+K_{\perp}+A_{\perp}))}, \\
  \omega_{[100],2} = \omega_{[010],2} &= 2\sqrt{(K+K_{\perp}+A_{\perp})(2J_{\perp}+K_{\perp}+A_{\perp})}.
\end{align*}
Note that the symmetric anisotropies have dropped out. For these wave-vectors we also see that
the effect of $K_{\perp}$ and $A_{\perp}$ cannot be distinguished, with only $K_{\perp}' \equiv K_{\perp}+A_{\perp}$ appearing. We see that, as expected, all zero modes have are gapped out when tetragonal distortions such as $J_{\perp} \neq J$, $K_{\perp} \neq K$ or $A_{\perp} \neq 0$ are included.

\subsubsection{Best Fit to Tetragonal Model}
Given the observed spectrum we can make reasonable inferences on which of these modes would correspond to which experimental features. First, the $\omega_{(001),1}$, $\omega_{(100),1}$ and $\omega_{(010),1}$ modes become gapless when $J_{\perp}=J$, $K'_{\perp}=K$. Following how these are expected acquire gaps in Fig.~\ref{supp:fig:swt-best-fit} for the $Z$ domain, we identify $\omega_{(001),1} \approx 0.65$~meV and $\omega_{(100),1}=\omega_{(010),1} \approx 0.9$~meV. The $\omega_{(001),2}$ mode would then correspond to the higher-lying mode at $(0,0,0)$ which was observed in ESR near $\omega_{(001),2} \approx 1.2\meV$. The final mode would then correspond to the high intensity mode near $\omega_{(010),2}=\omega_{(100),2}\approx 1.6 \meV$ at $(1,1,0)$, which also appears at $(1,0,0)$ and equivalents for this domain. These experimental constraints do not uniquely determine the four spatially anisotropic exchange parameters. In fact, this linear spin-wave model cannot reproduce all four energies due to satisfying the relation $\omega_{(001),1}^2+\omega_{(001),2}^2 = \omega_{(100),1}^2+\omega_{(100),2}^2$ which is not satisfied for the experimentally determined energies. 

\begin{figure}
    \includegraphics[width=\textwidth]{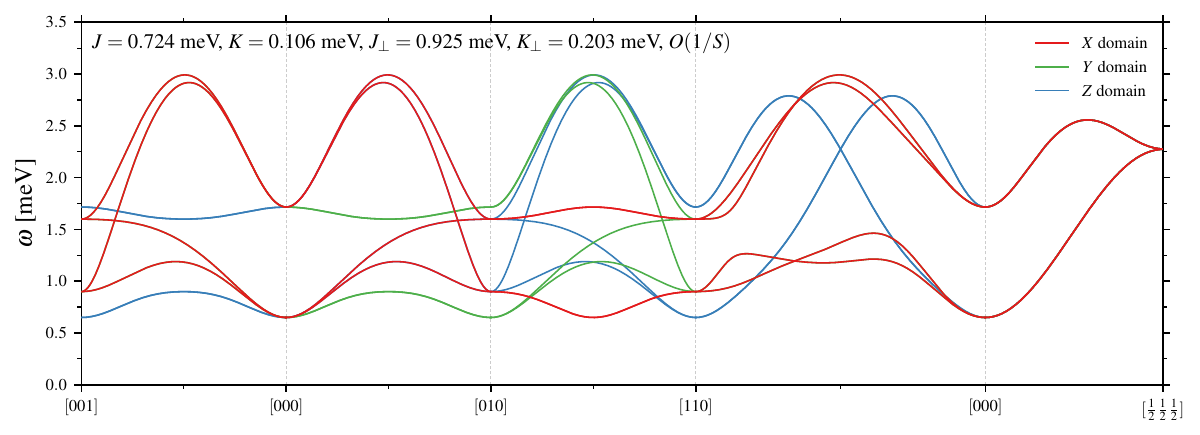}
    \caption{
    Linear  spin-wave spectrum of the nearest-neighbor model with tetragonal distortion [Eq.~(\ref{supp:eq:tetragonal-model})] in the type III antiferromagnet along a high-symmetry path 
    for parameters $J=0.724$ meV, $K=0.106$ meV, $J_{\perp}=0.935$ meV and $K_{\perp}=0.203$ meV with $\Gamma=\Gamma_{\perp}=A_{\perp}=0$. Three symmetry related domains are shown (with the tetragonal distortion following the domain direction). We see that the anisotropy introduces a large gap to all of the nodal lines and pseudo-Goldstone modes present in the undistorted case.
    \label{supp:fig:tet-best-fit}
    }
\end{figure}

To resolve this we take two steps: first, we only consider the two low lying modes ($0.65$~meV and $0.9$~meV) and the highest lying mode ($1.6$~meV).  To find a unique solution we additionally require that the Curie-Weiss constant $\Theta_{\rm CW} = -29.3$~K~\cite{reig-i-plessis:2020} is reproduced, with $(2J +J_{\perp}) + (2K+K_{\perp})/3 \approx 2.52$~meV.
This yields the solution
\begin{subequations}
\begin{align*}
  J &= 0.724 \meV, &
  K &= 0.106 \meV, \\
  J_{\perp} &= 0.935 \meV, &
  K_{\perp}' &= 0.203 \meV. 
\end{align*}
\end{subequations}
These parameters reproduce the three stated spin-wave energies as well as the Curie-Weiss temperature, but (necessarily) give an incorrect $\omega_{(001),2} \approx 1.71$~meV instead of $1.2$~meV for the excluded ESR-visible mode.
This represents a moderate increase in the Heisenberg coupling $J$ of about $29\%$ for the perpendicular plane $J_{\perp}/J=1.29$, but a near doubling of the effective Kitaev coupling $K_{\perp}'/K = 1.93$. Note that this could be partitioned arbitrarily into a smaller increase $K$, but a corresponding non-zero value of $A_{\perp}$. The spectrum of this best fit using linear spin-wave theory is shown in Fig.~\ref{supp:fig:tet-best-fit}.
\subsection{Estimating the Magneto-elastic Coupling}
\label{supp:sec:magneto-elastic-coupling}
As structural distortions due to magneto-elastic couplings can also (effectively) lower the symmetry and gap out the pseudo-Goldstone modes, it is worthwhile to estimate (at least roughly) the expected energy scale for such effects in \KIC{}. We will consider two approaches: first, we will estimate the change in exchange constants due to the observed tetragonal distortion below $T_{\rm N}$. Next we will directly estimate the distortion and induced exchanges based on more microscopic arguments. For both we find the effects of magneto-elastic coupling are far too small to account for the size of the gap observed in \KIC{}.

We follow the treatment of~\citet{liu:2019} that was developed to explain features of another iridate compound, Sr\tsub{2}IrO\tsub{4}, with similar on-site and exchange physics~\cite{porras:2019}. Consider a magneto-elastic coupling of the form
$$
\tilde{g}  \sum_{ij} \sum_{\mu\nu} \epsilon_{\mu\nu}\sum_{\mu'\nu'} C^{\mu\nu,\mu'\nu'}_{ij} S^{\mu'}_i S^{\nu'}_j
$$
where $\epsilon_{\mu\nu}$ is the (symmetric) strain, $\tilde{g}$ is the overall strength of
the magneto-elastic couplings and $C^{\mu\nu,\mu'\nu'}_{ij}$ represent dimensionless factors that form the correct spin quadrupoles out of the bond operators $ S^{\mu'}_i S^{\nu'}_j$ to couple to the $\epsilon_{\mu\nu}$ strain component. We expect $\tilde{g}$ to be decomposable into $\tilde{g} = \kappa g$ where $g$ is an atomic energy scale and $\kappa$ relates to the energy scale of exchange interactions. For the Ir\tsup{4+} we use an estimate~\cite{liu:2019} of $g \sim 5$ eV, while we assume $\kappa$ scales with the exchange $J$. Since the exchange energy scale in \KIC{} is a factor of $10^2$ smaller than in Sr\tsub{2}IrO\tsub{4}~\cite{jackeli:2009,liu:2019}, we will take $\kappa \sim 5 \cdot 10^{-5}$, thus yielding $\tilde{g} \sim 0.25$ meV.

The observed splitting of the structural Bragg peaks (see Fig.~\ref{fig:SM3}) below $T_{\rm N}$ implies a strain of roughly $\epsilon \sim 10^{-4}$. For this value of strain
the modification of the magnetic exchanges would then be of order $\tilde{g} \epsilon \sim 0.025$ $\mu$eV. This is negligible relative to all magnetic energy scales in \KIC{}. Roughly, one would expect for type I pseudo-Goldstone modes~\cite{rau:2018} that the induced gap would scale as $\Delta \sim \sqrt{JD}$ with $D \sim \tilde{g}\epsilon$. This yields a gap of $\sim 5$ $\mu$eV; as expected, far too small to account for the observed gap in \KIC{}.

Consider now the expected distortion due to $\tilde{g}$. This is given by 
$\epsilon \sim \tilde{g}/\mathcal{K}$ where $\mathcal{K}$ defines the elastic energy $\sim \frac{1}{2} \mathcal{K} \epsilon^2$. If we estimate that the elastic constants in \KIC{} are comparable to that in Sr\tsub{2}IrO\tsub{4}, then we would have an induced exchange $D \sim \tilde{g}^2/\mathcal{K}$ and thus gaps of size $\Delta \sim \sqrt{JD} = \tilde{g} \sqrt{J/\mathcal{K}}$. With $J$ and $\tilde{g}$ two orders of magnitude smaller in \KIC{}, this yields a factor of $10^3$ in the gap relative to the estimate of~\citet{liu:2019}, i.e. $\Delta \sim O(\mu{\rm eV})$ as above. One might expect a smaller elastic constant in \KIC{}, e.g. due to some evidence for proximity to a structural distortion~\cite{wang:2024}, but several orders of magnitude difference in $\mathcal{K}$ would be required to render the estimate for $\Delta$ significant. Note that for $\mathcal{K} \sim 100$ GPa yields an estimate of $\epsilon \sim 10^{-6}$ or $10^{-5}$; not too far from the observed $10^{-4}$ distortion.

\subsection{Dynamics in Paramagnetic Phase}
\label{supp:sec:paramagnetic}
Accessing dynamical properties in the paramagnetic regime when $T > T_N$ is difficult when quantum effects are included. We therefore adopt a classical approach for the spin dynamics in the regime, including renormalizations in the energy scale and intensities to partly mimic some quantum features.

We consider the Hamiltonian Eq.~(\ref{supp:eq:jkg}) augmented with second-neighbor exchange [Eq.~(\ref{supp:eq:j2})] in the classical limit where the spins are unit length vectors $|\vec{S}_i|^2=1$. We assume they follow conventional Landau-Lifshitz dynamics
\begin{equation}
    \label{supp:eq:lldyn}
    \frac{d\vec{S}_i}{dt} = -\vec{S}_i \times \frac{\del H}{\del \vec{S}_i},
\end{equation}
where $-\del H/\del \vec{S}_i \equiv \vec{B}_i$ is the (local) exchange field due to the neighboring spins. The initial conditions $\vec{S}_i(0)$ for this equation are drawn from a thermal distribution at temperature $T$ using Monte Carlo sampling. Once a sample of trajectories $\vec{S}_i(t)$ are obtained, the part of the dynamical structure factor relevant for inelastic neutron scattering is given by
$$
\mathcal{S}_{\rm cl}(\vec{k},\omega) = \sum_{\mu\nu} \left(\delta_{\mu\nu} - \hat{k}_\mu\hat{k}_\nu \right) \avg{\bar{S}^\mu_{\vec{k}}(\omega) S^{\nu}_{\vec{k}}(\omega)},
$$
where $\vec{S}_{\vec{k}}(\omega)$ is the Fourier transform of $\vec{S}_i(t)$ in both space and time. 
To account for our spins being $S=1/2$ we simply rescale frequencies by a factor of $S$, ensuring (for example) that in the low temperature limit the classical and quantum spin-wave frequencies agree. 

To partially account quantum effects we also multiply by an energy dependent correction factor~\cite{zhang:2019, scheie:2022}
$$
F_{\rm qu}(\omega) \equiv \beta\omega\left(1+n_B(\omega)\right),
$$
where $n_B(\omega) = 1/(e^{\beta\omega}-1)$ is the Bose factor.
This factor enforces the quantum version of the fluctuation-dissipation theorem relating positive and negative frequencies~\cite{chaikin:1995}, as $F_{\rm qu}(-\omega) = e^{-\beta \omega} F_{\rm qu}(\omega)$. 

Practically, we generate our thermal samples using Monte Carlo with single-site heat-bath~\cite{miyatake:1986} and over-relaxation~\cite{creutz:1987} updates, annealing down from high temperature $O(10J)$ to the temperature of interest and then thermalizing for the same number of sweeps. At the high temperatures of interest here only a small number of sweeps, typically $O(10^3)$, are necessary to reach equilibrium. 

For each sample we solve the coupled non-linear ordinary differential equations of Eq.~(\ref{supp:eq:lldyn}) using an adaptive fourth-order Runge-Kutta method~\cite{dormand:1980} as implemented in the \texttt{odeint} library~\cite{ahnert:2011}. Fourier transforms in both space and time, as implemented in the \texttt{FFTW} library~\cite{frigo:2005,bowman:2010}, are then used to obtain $\vec{S}_{\vec{k}}(\omega)$ and then compute the dynamical structure factor. At the temperatures of interest $O(10^2)$ samples are sufficient to reach convergence in both energy resolved and energy averaged quantities. The adaptive time stepping was performed with (absolute and relative) error tolerances of $10^{-8}$. The final trajectories were evaluated by interpolation on a grid to obtain a frequency spacing of $\Delta \omega = 0.04$ meV and a maximum frequency of $\omega_{\rm max} = 3$ meV, to allow comparison to the full measured spectrum.

For the simulations presented in the main text we used our best fit parameters, $J=0.74$ meV, $K=0.15$ meV and $\Gamma=0$, A small finite $J_2$ can be included to classically stabilize a type III order (see Sec.~\ref{supp:sec:j2-neel-temp}) though it does not significantly affect any of our results. To make meaningful comparisons to experiment, we performed simulations at $T=2T_N$ where $T_N$ is the (classical) N\'eel temperature to match the 6 K used experimentally. Each spectrum and slice is averaged in the same way as the experimental data, as described in more detail in Sec.~\ref{supp:sec:avg}.

\subsection{Effect of $J_2$ on N\'eel temperature}
\label{supp:sec:j2-neel-temp}
Monte Carlo simulations for the static properties of our best model fit (augmented with finite $J_2/J$) were carried out using parallel tempering Monte Carlo~\cite{hukushima:1996,swendsen:1986,marinari:1992}. Each temperature was annealed from the highest temperature in our grid then thermalized before samples were taken.  Single-site heat-bath~\cite{miyatake:1986} and over-relaxation~\cite{creutz:1987} updates were used throughout and proved sufficient to equilibriate our system when combined with the parallel tempering. Typical simulations were for systems of $L^3$ cubic cells ($N=4L^3$ spins) with $L=8,10,12$ and used $4\cdot 10^4$ sweeps for annealing and thermalization and a further $10^5$ sweeps for sample production. Each sweep consists of $N$ heat bath and $N$ over-relaxation updates followed by a sweep attempting swaps of (random) neighboring temperatures for each replica.

The classical transition temperature $T_N/J\sim 0.67$ was located via the position of the maximum in heat capacity and in the onset of the static structure factor at $(1, 0.5, 0)$ and equivalents. We confirmed that the $(1,0,0)$ type I order parameter remained small through the transition.  Note that with $J_2/J=0$ the transition remains near $T/J\sim 0.66$ but the order switches to type I due to the effects of order-by-thermal-disorder as has been established in the Heisenberg case~\cite{gvozdikova:2005}. While, small finite $J_2/J>0$ has little effect on $T_N/J$ the entire phase below $T_N$ switches to type III almost immediately. 

\subsection{Constraints on Size of $\Gamma$ Exchange}
\label{supp:sec:effect-of-gamma}
At the level of linear spin-waves the effect of finite $\Gamma$ is limited. For the spectra, slices and cuts shown in Fig.~2 of the main text the contributions to $\mat{A}_{\vec{k}}$ and $\mat{B}_{\vec{k}}$ that are $\propto \Gamma$ all vanish. We thus do not expect significant changes to our results if small $\Gamma \neq 0$ is included. 

To determine whether $\Gamma$ is not necessary to explain our data, we have examined $I(\vec{k},\omega)$ at wave-vector $\vec{k}=[0.5, 0.5 , 0.5]$ where $\Gamma$ would expected to contribute maximally at the level of linear spin-wave theory. We find no significant differences between our non-linear spin-waves with $\Gamma=0$ and the experimental data. Finally, we note that ab-initio calculations~\cite{khan:2019} estimate $\Gamma/J \approx 0.08$, small relative to $J$ and a factor of three smaller than our best fit $K/J = 0.2$.

Going beyond linear spin-waves a finite $\Gamma$ can induce
spontaneous magnon decay~\cite{zhitomirsky:2013} when the one- and two-magnon excitations overlap in energy. As discussed in Sec.~\ref{supp:sec:corrections-to-dyn-sf}, the two-magnon intensity would be expected to begin at 
 $2\Delta \sim 1.3 \meV$ where $\Delta \sim 0.65 \meV$ is the one-magnon gap. Any spontaneous decay would thus be limited to that regime. The two magnon density of states grows slowly starting from its minimum energy, further suppressing this decay channel. Given the resolution of the inelastic data presented here it is unlikely this relatively small level of decay could be resolved.

\subsection{Details of Comparison Between Theoretical and Experimental Dynamical Structure Factor}
\label{supp:sec:avg}
To perform a detailed comparison between the theoretical calculation and experimental data we must model not only $I(\vec{k},\omega)$, but also the finite energy resolution, averaging of the data in $\vec{k}$ and the magnetic form factor of Ir\textsuperscript{4+}.

\subsubsection{Energy Resolution}
Due to the finite energy resolution of the experimental measurement, we convolve our theoretical result with a Gaussian lineshape with width $\sigma_{\omega} = 0.07$ meV, chosen to match the peak widths observed in the individual $\vec{k}$ cuts [Main text, Fig. 2(g)]. The resulting intensity is then binned on the same grid as the experimental data with the bin width divided out. 

\subsubsection{Wave-vector Averaging}
To improve statistics the experimental spectrum [Fig.~2(a-b)], slices [Fig.~2(c-f)]
and cuts [Fig.~2(g)] presented in the main text have all been averaged over a window in wave-vector. 
\begin{itemize}
    \item \emph{Spectrum:} This integration window is $\pm 0.15$ r.l.u in the directions perpendicular to the path. In panel (a) this is integrated over wave-vectors $[1+\delta h,k,\delta l]$ for $-0.15 \leq \delta h, \delta l \leq 0.15$ and in panel (b) over $[\delta h,k,\delta l]$ over the same range.
    \item \emph{Slices:} This integration window is $\pm 0.2$ r.l.u in the direction perpendicular to the plane. In each panel the intensity has been averaged over wave-vectors $[h,k,\delta l]$ with $-0.2 \leq \delta l \leq +0.2$.
    \item \emph{Cuts:} This integration window is $\pm 0.1$ r.l.u each the direction. For a panel showing wave-vector $(h,k,l)$ the intensity has been averaged over wave-vectors $[h+\delta h,k+\delta k,l+\delta l]$ with $-0.1 \leq \delta h,\delta k, \delta l \leq +0.1$.
\end{itemize} 
In the theoretical calculations this averaging is carried out stochastically. For each point requiring averaging, we generate a uniform random wave-vector in the required range and calculate the intensity via sampling, with $O(10^2)$ samples sufficient to achieve reasonable convergence.

\subsubsection{Iridium Form Factor}
\begin{figure}
    \centering
    \includegraphics[width=0.6\textwidth]{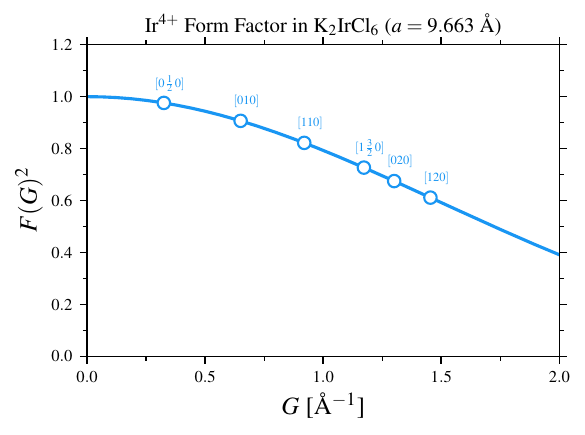}
    \caption{Ir\tsup{4+} magnetic form factor squared, with several wave-vectors relevant for \KIC{} shown. Radial integrals are obtained from~\cite{kobayashi:2011}.
    \label{supp:fig:ir-form-factor}
    } 
\end{figure}
The theoretical results must be multiplied by the magnetic form factor of Ir\textsuperscript{4+} to be meaningfully compared to the experimental data. We use the form factor calculated in~\citet{kobayashi:2011}, as has been used in previous neutron scattering studies of \KIC{}~\cite{aczel:2019}. We show the squared form factor in  Fig.~\ref{supp:fig:ir-form-factor}. We see the effect of the form factor can result in a reduction in intensity as high as 40\% in the range of wave-vectors of interest.

\subsubsection{Interpolation of Theoretical Results}
Due to the extensive averaging in the experimental data, we have adopted an interpolation procedure to calculate the linear and non-linear spin-wave energies and intensities at arbitrary points in the Brillouin zone. Interpolation of $\epsilon_{\vec{k},n}$ or $W_{\vec{k},n}$, while natural, does not correctly capture important features such as band crossings in the spectrum, as they are not smoothly varying as a function of wave-vector. 

Instead, we interpolate the full spin-wave dispersion matrix $\vec{M}_{\vec{k}}$ at the desired wave-vector and then use this interpolated matrix to calculate spin-wave energies and intensities. In linear spin-wave theory $\vec{M}_{\vec{k}}$ is an analytic function of $\vec{k}$ and thus can be interpolated effectively. In the self-consistent case, while not guaranteed, we expect the presence of a gap to render the corrections $\delta \mat{A}_{\vec{k}}$, $\delta \mat{B}_{\vec{k}}$ well-behaved (smooth) as a function of wave-vector. This procedure has the advantage of correctly capturing the dispersion and intensities near features like band crossings and pseudo-Goldstone modes. 

For all plots in the main text and supplemental materials we use a standard tricubic interpolation~\cite{lekien:2005} of the matrix elements of $\mat{M}_{\vec{k}}$. Due to our Fourier transform convention we have $\mat{M}_{\vec{k}} = \mat{M}_{\vec{k}+\vec{G}}$ for reciprocal lattice vectors $\vec{G}$ and thus the interpolation wraps appropriately through the periodic boundaries of our $\vec{k}$-grid. 

For our self-consistent spin-wave results we have used a finite system size of $N=4L^3$ with $L=12$ or $L=24$ to calculate the renormalized $\mat{M}^{\rm eff}_{\vec{k}}$ dispersion matrices. We have checked that no meaningful difference is visible between the $L=12$ and $L=24$ interpolated results, verifying that the scheme has converged in system size and that the interpolation procedure is effective. For the linear spin-wave results we use the same scheme, but with $L=64$.
\putbib
\end{bibunit}

\end{document}